 \documentclass[apj,numberedappendix]{emulateapj}

 \usepackage{color}
\usepackage{amsmath}
\usepackage{enumerate}
\usepackage{mathrsfs}
\usepackage{natbib}
\usepackage{hyperref}
\usepackage{ulem}

 \usepackage{isomath}
 \usepackage{bbold}

\def\simlt{\lower.5ex\hbox{$\; \buildrel < \over \sim \;$}}
\def\simgt{\lower.5ex\hbox{$\; \buildrel > \over \sim \;$}}

\def\beq{\begin{equation}}
\def\eeq{\end{equation}}
\def\ba{\begin{eqnarray}}
\def\ea{\end{eqnarray}}
\def\bB{{\,\mathbf B}}
\def\bE{{\,\mathbf E}}

\def\bj{{\,\mathbf j}}
\def\bv{{\,\mathbf v}}

 \def\XTE{XTE~J1810-197}

\def\Sect{{\rm Section}} 
\def\Sects{{\rm Sections}} 

\def\Eq{Equation}
\def\Eqs{Equations}

\def\spl{s}
\def\dspl{\dot{s}}

\def\sigcr{\sigma_{\rm cr}}

\def\tohm{t_{\rm ohm}}

\def\dsav{\bar{\dot{s}}}

\def\zm{z_{\rm melt}}
\def\Tm{T_{\rm melt}}
\def\Fh{F_h}
\def\Fs{F_s}
\def\Emag{E_{\rm mag}}

\def\vH{v_{\rm H}}
\def\bvH{{\mathbf v}_{\rm H}}

\def\dqohm{\dot{q}_{\rm ohm}}
\def\eff{\epsilon}

\def\tc{t_c}
\def\cond{\tilde{\sigma}}

\def\Tc{T_{\rm core}}
\def\Eth{E_{\rm th}}
\def\thB{\Theta_B}

\def\Rpn{{\cal R}_{pn}}
\def\Rlambda{{\cal R}_\lambda}
\def\RD{{\cal R}_D}
\def\RM{{\cal R}_M}
\def\tpn{\tau_{pn}}
\def\dmu{\Delta\mu}

\def\rhonuc{\rho_{\rm nuc}}
\def\Tcrit{T_{\rm crit}}

\def\Erad{E_{\rm rad}}
\def\dB{\Delta B}
\def\ddB{\delta B}
\def\dqh{\dot{q}_h}
\def\dqnu{\dot{q}_\nu}
\def\Tbal{T_{\rm bal}}
\def\tdiss{t_{\rm diss}}
\def\Leff{L_1}
 \def\kk{\mathbb{k}}
 \def\sigSB{\sigma_{\rm SB}}
 \def\Eaft{E_{\rm aft}}
 \def\Eprompt{E_{\rm prompt}}
 \def\fw{f_{\rm wave}}
 \def\Ls{L_\star}
 
 \def\tlam{t_{\lambda}}
 \def\sSB{\sigma_{\rm SB}}
 \def\Lum{\mathscr L}

\newbox\grsign \setbox\grsign=\hbox{$>$} \newdimen\grdimen \grdimen=\ht\grsign
\newbox\simlessbox \newbox\simgreatbox \newbox\simpropbox
\setbox\simgreatbox=\hbox{\raise.5ex\hbox{$>$}\llap
     {\lower.5ex\hbox{$\sim$}}}\ht1=\grdimen\dp1=0pt
\setbox\simlessbox=\hbox{\raise.5ex\hbox{$<$}\llap
     {\lower.5ex\hbox{$\sim$}}}\ht2=\grdimen\dp2=0pt
\setbox\simpropbox=\hbox{\raise.5ex\hbox{$\propto$}\llap
     {\lower.5ex\hbox{$\sim$}}}\ht2=\grdimen\dp2=0pt
\def\simgt{\mathrel{\copy\simgreatbox}}
\def\simlt{\mathrel{\copy\simlessbox}}

\defcitealias{1996ApJ...473..322T}{TD96}

\begin{document}

\title{
Magnetar heating
}

 \author{
 Andrei M. Beloborodov
 and {Xinyu Li}
 }
 \affil{
 Physics Department and Columbia Astrophysics Laboratory,
Columbia University, 538  West 120th Street New York, NY 10027;
 amb@phys.columbia.edu}

\begin{abstract}
We examine four candidate mechanisms that could explain the high surface 
temperatures of magnetars.\\
(1) Heat flux from the liquid core heated by ambipolar diffusion.
It could sustain the observed surface luminosity $\Lum_s\approx 10^{35}$~erg~s$^{-1}$
if core heating offsets neutrino cooling at a temperature $\Tc>6\times 10^8$~K.
This scenario is viable if the core magnetic field exceeds $10^{16}$~G and the
heat-blanketing envelope of the magnetar
has a light element composition. We find however that the lifetime of such a hot core 
should be shorter than the typical observed lifetime of magnetars. \\
(2) Mechanical dissipation in the solid crust. This 
heating can be quasi-steady, powered by gradual 
(or frequent) crustal yielding to magnetic stresses. We show that it
obeys a strong upper limit. As long as the crustal stresses are fostered by the field 
evolution in the core or Hall drift in the crust, mechanical heating 
is insufficient to sustain persistent
$\Lum_s\approx 10^{35}$~erg~s$^{-1}$.
The surface luminosity 
is increased in an alternative scenario of mechanical deformations
triggered by external magnetospheric flares.\\
(3) Ohmic dissipation in the crust, in volume or current sheets. 
This mechanism is inefficient because of the high conductivity of the crust. 
Only extreme magnetic configurations with crustal fields $B>10^{16}$~G varying on 
a 100 meter scale could provide $\Lum_s\approx 10^{35}$~erg~s$^{-1}$.
\\
(4) Bombardment of the stellar surface by particles accelerated in the magnetosphere.
This mechanism produces 
hot spots on magnetars.
Observations of transient magnetars
show evidence for external heating.
\end{abstract}

\bigskip



\section{Introduction}

Heat stored in neutron stars after their birth is gradually lost to neutrino emission 
and surface radiation. As a result, a kyr-old neutron star is expected to have an internal 
temperature $T\approx 10^8$~K and a surface temperature $T_s\approx 10^6$~K 
\citep{2004ARA&A..42..169Y}.
This expectation is violated by magnetars, a special class of neutron stars with 
ultrastrong magnetic fields, $B\sim 10^{14}-10^{16}$~G. 
The ages of observed magnetars are $\sim 1-10$~kyr
and their persistent surface temperatures reach $5\times 10^6$~K, making them much 
more luminous than ordinary, passively cooling, neutron stars of the same age
(e.g. \citealp{2013MNRAS.434..123V}).
Persistent active magnetars show a remarkably narrow range of surface luminosities
around $\Lum_s\approx 10^{35}$~erg~s$^{-1}$ \citep{2006ApJ...650.1070D}. For
a neutron star of radius $R\approx 10-13$~km, this luminosity
corresponds to effective surface temperature $T_s\approx 4\times 10^6$~K,
which is consistent with the temperatures estimated from the shape 
of the observed soft X-ray emission.\footnote{Gravitational redshift reduces the 
     observed temperature by the factor of 
     $(1-2GM/c^2R)^{1/2}\approx 0.8$. On the other hand, radiation emerging from the 
     magnetar atmosphere is not exactly Planckian, which tends to somewhat increase the 
     observed temperature.
     In contrast to a normal blackbody, magnetar surface radiation is dominated by 
     one of the two polarization states (e.g. \citealp{2006RPPh...69.2631H}).}

By definition of magnetars, their luminosities are fed by magnetic energy stored in 
the neutron star \citep{1992ApJ...392L...9D,1992AcA....42..145P}. How can 
magnetic energy be converted to heat?

(1) One dissipative process is provided by ambipolar diffusion of the magnetized 
electron-proton fluid through the liquid neutron core
\citep[][hereafter TD96]{1992ApJ...395..250G,1996ApJ...473..322T}.
The rate of this process scales as $B^2$, which suggests its efficiency
in magnetars. Ambipolar diffusion could keep the core hot for some time,
and the heat flux from the core could sustain the observed surface temperature.
The challenge faced by this scenario is the enormous neutrino cooling that hinders
the heating of the core.

(2) Strong magnetic stresses deform the solid crust beyond the elastic limit, resulting 
in mechanical dissipation. Mechanical heating was envisioned in the starquake
picture of magnetar activity \citep{1995MNRAS.275..255T,1996ApJ...473..322T}; 
its more plausible version is a plastic flow
\citep{2002ApJ...574..332T,2003ApJ...595..342J,2014ApJ...794L..24B}. 
Mechanical heating can only occur in the solid crust below the melted ocean,
at depths $z\simgt 100$~m below the stellar surface.

(3) Magnetic fields in neutron stars gradually decay due to ohmic dissipation.
This mechanism is usually considered to be inefficient on the kyr timescales of interest,
because of a high electric conductivity of the crust (and a huge conductivity of the core).
Ohmic heating could become important in the presence of strong gradients of the 
magnetic field, which are sustained by strong electric currents. It was proposed that 
ohmic dissipation is assisted by the Hall drift, which can transport magnetic energy to 
the shallow subsurface layers \citep{1988MNRAS.233..875J}, where conductivity is 
lowest, and could develop a ``Hall cascade'' \citep{1992ApJ...395..250G}.
In addition, it was proposed that deformations of the crust by the magnetic stresses
could create current sheets where strong localized heating could
occur \citep{2001ApJ...561..980T,2002ApJ...580L..69L}. 

(4) The bombardment of the magnetar surface by magnetospheric 
particles results in its external heating. Evidence for high-energy particles is 
provided by persistent nonthermal emission from magnetars, which 
is associated with continual electron-positron discharge in 
the twisted magnetosphere \citep{2007ApJ...657..967B,2013ApJ...777..114B}. 

In this paper we examine the efficiencies of the heating mechanisms (1)-(4)
using simple estimates and illustrating with sample numerical models.


\section{Cooling of a hot core}

The heat capacity of a core with non-superfluid neutrons  determines the maximum 
thermal energy that could be stored in a neutron star (e.g.  \citealp{2004ApJS..155..623P}),
\beq
   \Eth\sim 10^{48} T_9^2  {\rm ~erg}.
\eeq
Without heating, most of $\Eth$ is lost to neutrino emission 
on a timescale shorter than the typical magnetar age $t\sim 10^{11}$~s,
and the core temperature decreases to $\Tc\sim 10^8$~K while its 
surface luminosity $\Lum_s$ drops well below $10^{35}$~erg~s$^{-1}$ 
\citep{2004ARA&A..42..169Y,2009ASSL..357..247P}.

In this section, we discuss what core temperature would be sufficient to sustain the 
observed $\Lum_s$ of active magnetars. Then we estimate the required heating that 
must offset the neutrino cooling to keep the core hot. Section~\ref{ambipolar} will address how the 
high temperature could be sustained by ambipolar diffusion.

\subsection{Core temperature capable of sustaining $\Lum_s$}

The surface luminosity of persistent magnetars $\Lum_s\approx 10^{35}$~erg~s$^{-1}$
approximately corresponds to the 
{\it average} surface flux
\beq
   F_s=\sSB T_s^4=\frac{\Lum_s}{A}
   =10^{22} A_{13}^{-1} {\rm ~erg~s}^{-1}~{\rm cm}^{-2},
\eeq
where $\sSB\approx 5.67\times 10^{-5}$~erg~cm$^{-2}$~s$^{-1}$~K$^{-4}$ is the
Stephan-Boltzmann constant, $T_s\approx 4\times 10^6$~K is
the effective surface temperature, and $A$ is the emission area, which may be 
smaller than the stellar surface area 
$4\pi R^2\approx 1.5\times 10^{13}$~cm$^2$. 

Such a high $T_s$ can be sustained if the interior temperature is comparable to 
$10^9$~K \citep{2004ARA&A..42..169Y}.
The interior region here includes 
not only the core ($\rho>1.4\times 10^{14}$~g~cm$^{-3}$) but also the lower 
crust ($\rho\gg 10^{11}$~g~cm$^{-3}$);
this region is nearly isothermal 
due to its high thermal conductivity.
A strong temperature gradient is sustained in the blanketing envelope
in the upper crust, especially where $\rho<10^{9}$~g~cm$^{-3}$,
because this region has a lower thermal conductivity.
A steady heat flux $F_s$ is established on the timescale of heat 
conduction across the crust, $\tc\sim 1-10$~yr.

The relation between $\Tc$ and $T_s$ depends on the strength of the magnetic field 
$\bB$
in the blanketing envelope and its angle with respect to the radial 
direction, $\thB$, because both affect heat conduction \citep{1999A&A...351..787P}. A strong radial magnetic field ($\thB=0$)
increases the heat flow to the surface. This is the result of Landau quantization of 
electron motion in the envelope (electrons can only move along $\bB$ at low densities 
where the electron Fermi energy is below 
the Landau energy $\hbar\omega_B$).
In contrast, a horizontal field ($\thB=\pi/2$) impedes the heat flow by the factor of
$(\tau\omega_B)^{-2}\ll 1$, where $\tau$ is the collisional free path time of electrons.

The $\Tc$-$T_s$ relation also depends on the chemical composition, which must be 
iron in the lower envelope $\rho\simgt 10^9$~g~cm$^{-3}$ 
but may be lighter elements in the 
surface layers.
\citet{2003ApJ...594..404P} calculated the $\Tc$-$T_s$ relation and gave its analytical approximation
 for various $B$ and $\thB$, for both iron and light element envelopes.
Their calculations assumed that neutrino cooling of the envelope is negligible
and the envelope has a gaseous atmosphere. The latter assumption may be invalid,
as the magnetar surface is likely condensed
\citep{2006PhRvA..74f2508M}, although only 
approximate calculations are available for the phase transition to the condensed state.

\citet{2007Ap&SS.308..353P} included neutrino cooling 
and studied heat conduction in stars with condensed surfaces.
They found that replacing the atmosphere with a condensed surface
weakly affects the $\Tc$-$T_s$ relation, and
neutrino losses in the envelope become important when $\Tc\simgt 10^9$~K.
The losses effectively impose a ceiling for the surface luminosity: $\Lum_s$ reaches 
its maximum value $\sim 10^{35}$~erg~s$^{-1}$ when $\Tc\sim 10^9$~K and does 
not respond to further increase of $\Tc$, because the heat flux is lost to neutrino 
emission on its way through the crust.
Below this ceiling the $\Tc$-$T_s$ relation from \citet{2003ApJ...594..404P},
with neglected neutrino losses, may be used.

\begin{figure}[t]
\begin{tabular}{c}
\includegraphics[width=0.5\textwidth]{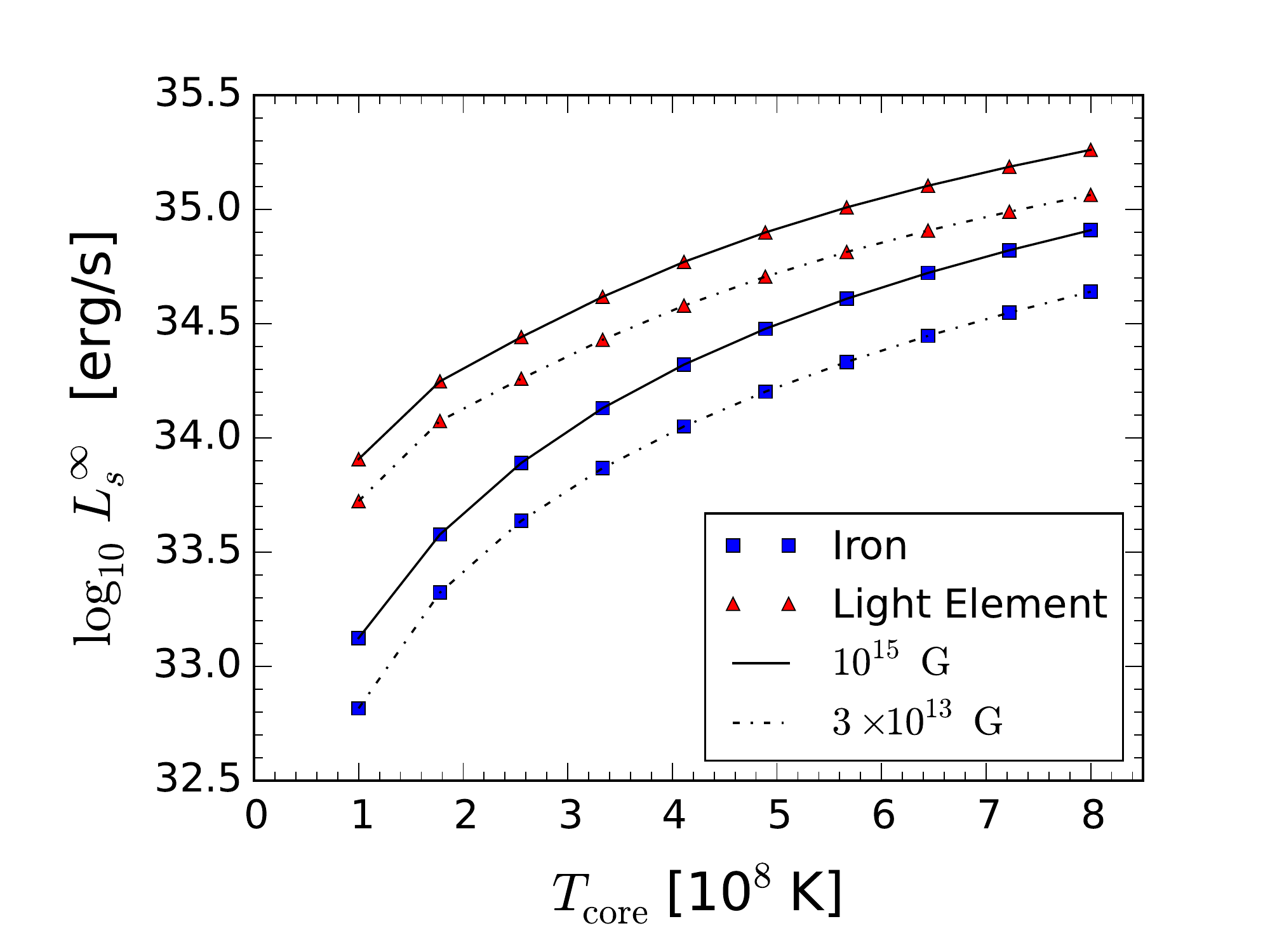} 
\end{tabular}
\caption{Surface luminosity emitted by a neutron star with a hot core, as 
observed at infinity. Each symbol shows a calculated model of steady heat 
transfer from the core to the stellar surface. The star is assumed to have 
a dipole magnetic field near the surface, in the heat blanketing envelope.
Two cases are considered: the iron envelope and the maximal light element 
envelope, which is called ``fully accreted" in \citet{2003ApJ...594..404P}. 
The luminosity is shown for two values of the polar magnetic 
field: $B_p=3\times 10^{13}$~G and a more typical for magnetars $B_p=10^{15}$~G.
As $\Tc$ approaches $10^9$~K, $\Lum_s^\infty$ approaches the ceiling imposed by 
neutrino cooling \citep{2007Ap&SS.308..353P}; heating the core to higher 
temperatures would not significantly increase the surface luminosity. }
 \label{fig_Tc}
 \end{figure}

The dependence of the surface luminosity on $\Tc$ is shown in Figure~\ref{fig_Tc} 
for a neutron star of mass $M=1.4M_\odot$ and radius $R=11.7$~km.
The envelope is assumed to have an approximately dipole magnetic field $\bB$;
then the angle between $\bB$ and the radial direction is
\beq
  \tan \thB=\frac{\sin\theta}{2\cos\theta},
\eeq 
where co-latitude $\theta$ is measured from the magnetic pole
and general relativistic corrections have been neglected.
The surface luminosity of the star is given by
\beq
  \Lum_s=4\pi R^2\int_{0}^{1} F_s(\thB,B)\,d\cos\theta,
\eeq
and the observed luminosity at infinity $\Lum_s^\infty=(1-2GM/c^2R)\Lum_s$ is reduced 
by the factor of 1.5. One can see 
from Figure~\ref{fig_Tc} that the strong magnetic field assists heat conduction to the surface, however in any case
a high core temperature is required to sustain 
$\Lum_s^\infty=10^{35}$~erg~s$^{-1}$.
In particular, $\Tc\approx 10^9$~K is required if the star 
has an iron envelope.
In the case of the maximum light element envelope,
the required $\Tc$ is reduced to $\approx 6\times 10^8$~K
(see also Figure~4 in \citealp{2009MNRAS.395.2257K}).

\subsection{Neutrino cooling of the core}

The high $\Tc\simgt 6\times 10^8$~K implies a high cooling rate due to neutrino 
emission.
Direct urca cooling (hereafter Durca) can provide a huge sink of energy 
in the center of the core,
\beq
   \dot{q}_\nu^D\sim 10^{27}\,T_9^6\, \RD {\rm~erg~s}^{-1}{\rm cm}^{-3} 
   \:\:\: (\rho\simgt 10^{15} {\rm~g~cm}^{-3}),
\eeq
where $\RD\leq 1$ is a suppression factor that appears in the presence of 
superfluidity \citep{2001PhR...354....1Y}.  No reasonable heating mechanism can
compete with Durca cooling at temperatures $\Tc\sim 10^9$~K.
However, it is activated only if the separation between the Fermi levels of protons and 
neutrons is sufficiently small, which requires a minimum density comparable to 
$10^{15}$~g~cm$^{-3}$, and hence a minimum mass of the neutron star 
\citep{1991PhRvL..66.2701L}. The exact threshold mass for the onset of Durca, 
$M_D$, depends on the equation of state of the core matter 
\citep{1998PhRvC..58.1804A,2011PhRvC..84f2802C,2013A&A...560A..48P}
and can significantly exceed the canonical neutron star mass $M=1.4M_\sun$.

Stars with masses $M<M_D$
do not activate Durca, and the cooling occurs with a lower rate due to  
the modified urca reactions (hereafter Murca), which involve a spectator nucleon
taking the excess momentum.
Murca occurs everywhere in the core with the cooling rate given by 
(\citealp{1979ApJ...232..541F}),
\beq
\label{eq:Murca}
  \dot{q}_{\nu}^M\sim  7\times 10^{20}
  \,T_9^8 \left(\frac{\rho}{\rhonuc}\right)^{2/3} \RM
    {\rm ~erg~s}^{-1}{\rm ~cm}^{-3},
\eeq
where $\rhonuc=2.8\times 10^{14}$~g~cm$^{-3}$ is the nuclear saturation density.
With the onset of proton or neutron superfluidity the Murca rate is suppressed by the 
factor $\RM<1$, and the main cooling process becomes``Cooper pair cooling'' ---  
neutrino emission that accompanies the formation and breaking of Cooper 
pairs \citep{1976ApJ...205..541F,2006MNRAS.365.1300K,2009ApJ...707.1131P}.
Its rate is given by
\beq
   \dqnu^{CP}\sim 10^{21} \left(\frac{\rho}{\rhonuc}\right)^{1/3} T_9^7 
    \; f\left(\frac{\Tc}{\Tcrit}\right) {\rm ~erg~s}^{-1}{\rm ~cm}^{-3}, 
\eeq
where 
$\Tcrit$ is the critical temperature for the transition to superfluidity, and
the numerical factor $f(\Tc/\Tcrit)$ describes 
the temperature dependence of the Cooper pair cooling;
$f=0$ at $\Tc>\Tcrit$, $f$ steeply reaches a 
maximum at $\Tc\approx 0.8\Tcrit$ and steeply declines at $\Tc<0.5 \Tcrit$.

\begin{figure}[t]
\hspace*{-8mm}
\begin{tabular}{c}
\includegraphics[width=0.55\textwidth]{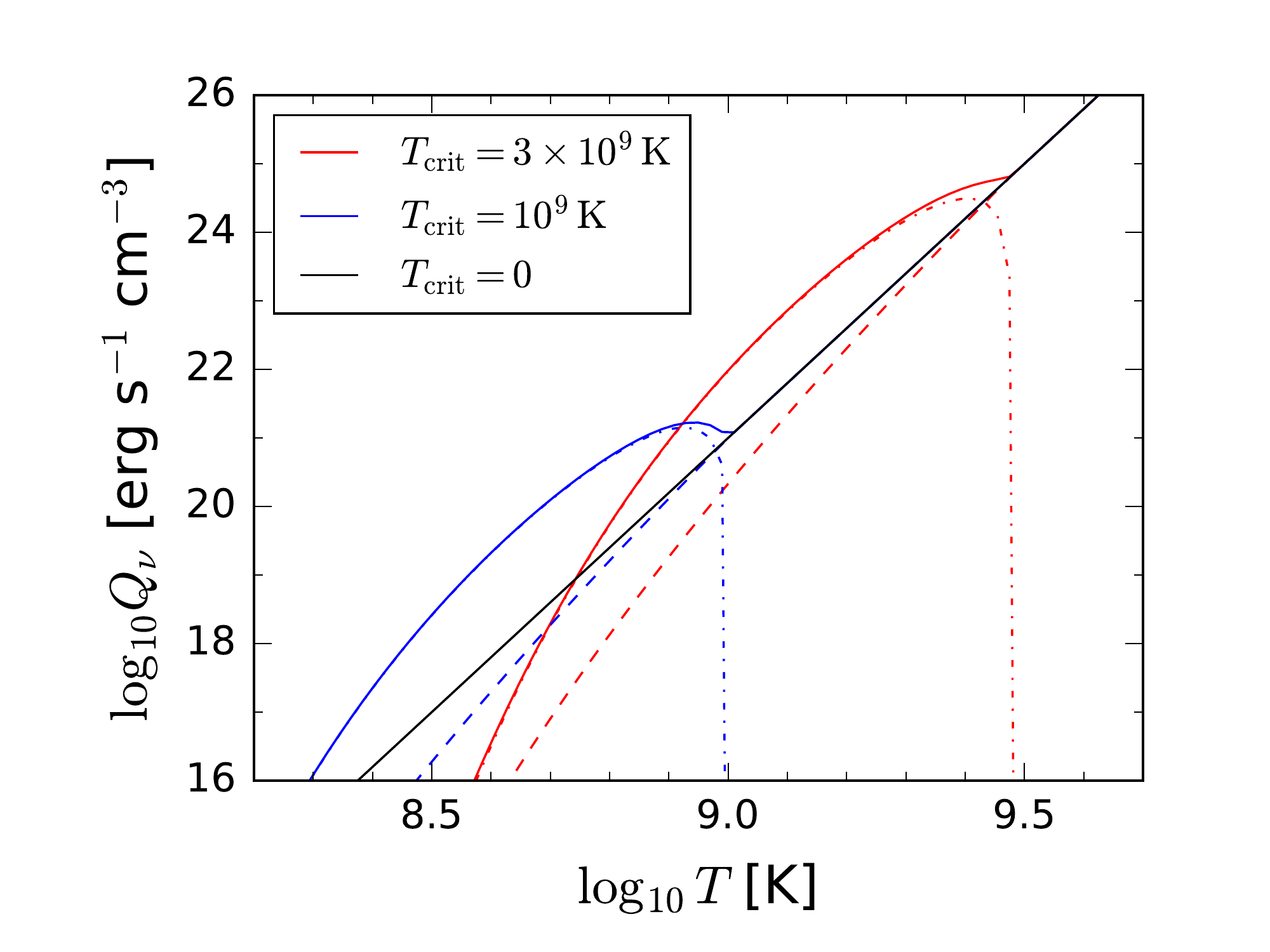} 
\end{tabular}
\caption{Neutrino cooling rate as a function of temperature in the core at density 
$\rhonuc=2.8\times 10^{14}$~g~cm$^{-3}$. Black curve shows Murca cooling 
assuming no superfluidity ($\Tcrit<10^8$~K). Colored curves show the cooling 
of matter with non-superfluid protons and superfluid neutrons, for two cases: 
$\Tcrit=10^9$~K (blue curves) and $\Tcrit=3\times 10^9$~K (red curves). 
Dashed curve shows the Murca contribution and dash-dotted curve shows the 
Cooper pair contribution; the net cooling rate is shown by the solid curve.
The triplet-state neutron pairing is assumed 
(model B in \citealp{2001PhR...354....1Y}).
 }
 \label{fig_qv}
 \end{figure}

The putative internal magnetic fields $B\simgt 10^{16}$~G are sufficiently strong to quench 
Cooper pairing of protons in most of the core volume, except perhaps its center
\citep{1969Natur.224..673B}. Therefore, we assume that protons are normal, not 
superfluid. The onset of neutron superfluidity is theoretically expected in the core at a 
temperature $\Tcrit\sim 10^8-10^9$~K (see e.g. Figure~5 in \citealp{2015SSRv..191..239P}).
There is some observational evidence for this transition from the observed surface 
temperatures of isolated neutron stars, however this is not settled 
and there remains a significant uncertainty in $\Tcrit$
\citep{2004ARA&A..42..169Y,2009ApJ...707.1131P,2012MNRAS.422.2632H,2015PhRvC..91a5806H}.

Based on the detailed calculations summarized by \citet{2001PhR...354....1Y}, 
Figure~\ref{fig_qv} shows the temperature dependence of the neutrino cooling rate 
$\dqnu=\dqnu^M+\dqnu^{CP}$ at the characteristic nuclear density 
$\rhonuc=2.8\times 10^{14}$~g~cm$^{-3}$.
One can see that the onset of superfluidity {\it increases} neutrino cooling in the 
temperature range of main interest $T>6\times 10^8$~K. 

Using the conservative (non-superfluid Murca) cooling rate one can estimate the minimum
neutrino luminosity as $\Lum_\nu=V_c\, \dot{q}_\nu\sim 10^{39}\, T_9^8$~erg~s$^{-1}$,
where $V_c\sim 10^{18}$~cm$^{3}$ is the volume of the core.
Sustaining a hot core over the typical magnetar age $t\sim 10^{11}$~s
requires deposition of energy 
\beq
\label{eq:Ecool}
   E\approx \Lum_\nu t \sim 10^{50}\,T_9^8\, {\rm~erg}.
\eeq
This rough, conservative estimate should be compared with the magnetic energy stored 
in the neutron star,
\beq
\label{eq:Emag_int}
  \Emag\approx \frac{4\pi}{3}\,R^3\, \frac{B^2}{8\pi} \sim 10^{49}\,B_{16}^2 {\rm ~erg}. 
\eeq
Comparison of \Eqs~(\ref{eq:Ecool}) and (\ref{eq:Emag_int}) shows that
internal magnetic fields $B\sim 10^{16}$~G are required to provide energy 
for interesting heating capable of sustaining  
$\Tc>6\times 10^8$~K and surface luminosity $\Lum_s\sim 10^{35}$~erg~s$^{-1}$.

The reservoir of magnetic energy $\Emag\sim 10^{48}-10^{49}$~erg is consistent with the
observed radiation output of magnetars. 
After three decades of observations of SGR~1806-20, a giant flare 
radiated $\sim 2\times 10^{46}$~erg \citep{2005Natur.434.1107P}, 
and the total energy output over the magnetar lifetime is likely to approach
$\Erad\sim 10^{48}$~erg. Assuming a reasonable 
efficiency $\Erad/\Emag\simlt 0.1$, the inferred magnetic energy is 
$\Emag\sim 10^{49}$~erg.


\section{Magnetic dissipation in the core}\label{ambipolar}

\subsection{Ambipolar drift}

The main
process capable of dissipating magnetic energy in the core is ambipolar diffusion 
(\citealt{1992ApJ...395..250G}, \citetalias{1996ApJ...473..322T}).
Ambipolar drift is the motion of the e-p plasma through the (approximately static) 
neutron fluid. 
The drift is driven by the  Lorentz force $\bj\times\bB/c=(\nabla\times\bB)\times\bB/4\pi$ 
and tends to relieve the magnetic stresses that drive it. Below we summarize
the standard description of ambipolar diffusion in a neutron star core and  then examine 
its role in magnetars.

The drift is opposed by two factors: pressure perturbations it 
induces and friction against the neutron fluid. Friction results from nuclear
collisions between protons and neutrons \citep{1990SvAL...16...86Y,2001A&A...374..151B};
electron-neutron collisions are negligible. The rate of p-n collisions per proton is given by
\beq
    \tpn^{-1}\approx 5\times 10^{18}\,T_9^2
    \left(\frac{\rho}{\rho_{\rm nuc}}\right)^{-1/3}\Rpn \; {\rm s}^{-1},
\eeq
where $\rhonuc\approx 2.8\times 10^{14}$~g~cm$^{-3}$, and $\Rpn=1$ if both 
protons and neutrons are non-superfluid. 
In the presence of superfluidity, $\Rpn<1$ describes the strong (asymptotically exponential) 
suppression of the collision rate \citep{2001A&A...374..151B}.

Pressure perturbations 
are induced if $\nabla\cdot(n_e \bv)\neq 0$, where $n_e=n_p$ is the electron/proton
number density and $\bv$ is the proton drift velocity. Such a ``compressive'' drift generates 
a change in $n_e$, and hence changes the electron and proton pressures, which 
are related to the chemical potentials $\mu_e$ and $\mu_p$ (Fermi energy levels). 
The resulting pressure gradient may be written as $-n_e\nabla(\dmu)$ 
where
\beq
   \dmu=\mu_e+\mu_p-\mu_n,
\eeq
which also describes a local deviation from chemical $\beta$-equilibrium 
$e,p\leftrightarrow n$.\footnote{For simplicity, our discussion here assumes
     the $n,p,e$ composition of the core. A more detailed model will need to include the 
     muon component that appears where the electron chemical potential exceeds the 
     muon rest-mass energy.}
The chemical potentials $\mu_e$, $\mu_p$, $\mu_n$ include the rest-mass 
energies of the species.

The pressure perturbation $\Delta P\sim n_e\dmu$ cannot exceed the magnetic 
stresses that drive the compressive drift ---  the drift is chocked when 
$n_e|\Delta\mu|\sim B^2/8\pi$. For $B<10^{17}$~G, the magnetic stresses
are small compared with the hydrostatic pressure in the core. Therefore, possible 
deviations $\dmu$ are small compared with the neutron chemical potential 
$\tilde{\mu}_n=\mu_n-m_nc^2$. 
The latter may be approximated as $\tilde{\mu}_n\approx 100\,(\rho/\rhonuc)^{2/3}$~MeV with a moderate accuracy of tens of percent, depending on the core equation of state.
Note that $\tilde{\mu}_n\gg (m_n-m_p-m_e)c^2$.
When evaluating quantities weakly affected by the small $\dmu$, such as 
plasma density $n_e$, one can use the approximate chemical balance 
$\tilde{\mu}_e+\tilde{\mu}_p\approx \tilde{\mu}_n$,  
where the chemical potentials with tilde do not include the rest-mass energies.
Note also that $\mu_e\approx\tilde{\mu}_e\gg \tilde{\mu}_p$,
because the degenerate electrons are ultra-relativistic and the 
degenerate protons are non-relativistic (while their number densities are equal).
Therefore the approximate equilibrium implies $\tilde{\mu}_e\approx\tilde{\mu}_n$.

The equation of ambipolar diffusion driven by the Lorentz force and opposed by p-n friction and pressure gradients reads \citep{1992ApJ...395..250G},
\beq
\label{eq:fbalance}
   \frac{(\nabla\times\bB)\times\bB}{4\pi}=n_e\nabla(\dmu) +\frac{n_e m_p^\star \bv}{\tpn},
\eeq
where $v$ is the proton velocity and $m_p^\star\approx 10^{-24}$~g is the effective 
proton mass. This equation takes into account that the drift is slow and one can neglect 
the $dv/dt$ term in the dynamic equation, i.e. the force balance is satisfied.

The charged-current weak interactions (in particular the Murca reactions) 
tend to restore $\beta$-equilibrium, i.e. to erase $\Delta\mu$.
The reaction rate may be written as $\dot{n}_e=-\lambda|\Delta\mu|$,
where $\lambda$ is related to the compressibility of the plasma \citep{1989PhRvD..39.3804S}. 
The low ``ceiling'' $|\Delta\mu|\ll \mu_e$ implies that significant 
compression or expansion can only proceed as allowed by the 
Murca reactions, i.e. there is an approximate balance,
\beq
\label{eq:compr}
  \nabla\cdot(n_e\bv)\approx -\lambda\,\dmu, 
  \qquad \left|\frac{\partial n_e}{\partial t}\right| \ll \left|\lambda\,\dmu\right|.
\eeq 
The value of $\lambda$ is given by
\beq
\label{eq:lambda}
   \lambda\approx 5\times 10^{33}\,T_9^6\left(\frac{\rho}{\rhonuc}\right)^{2/3} 
      H\,\Rlambda
      {\rm ~erg}^{-1}{\rm ~cm}^{-3}{\rm ~s}^{-1}.
\eeq
This expression takes into account the possible suppression of $\lambda$ due to
neutron superfluidity (factor $\Rlambda\leq 1$) and the enhancement due to the 
deviation from $\beta$-equilibrium (factor $H\geq 1$), see \citet{2001PhR...354....1Y}. 
The $H$-factor is significantly above unity when $\xi\equiv |\Delta\mu|/kT\gg 1$,
\begin{eqnarray}
   H(\xi)=\left\{ \begin{array}{ll}
   1                 & \quad \xi\ll 1 \\
   (0.11\xi)^6  & \quad \xi\gg 10.
                       \end{array} 
              \right.
\end{eqnarray}
In the regime $\xi\gg 10$ the Murca rate is independent of temperature. 
An explicit analytical expression for $H(\xi)$ is given by \citet{1995ApJ...442..749R}
and Appendix~\ref{app}. The factor $\Rlambda(T/\Tcrit)$ 
was calculated by \citet{2001A&A...372..130H}, where $\Tcrit$ is the temperature of
the superfluid transition (it appears that they mislabeled the curves in their Figure~2). 

The basic picture of ambipolar diffusion may be summarized as follows.
Let $L$ be a characteristic scale of the field variation $\dB$. Estimating 
$(\nabla\times\bB)\times\bB\sim B\,\dB /L$, $\nabla(\dmu)\sim\dmu/L$, and 
$\nabla\cdot(n_e\bv)\sim n_e v/L$, one finds 
\beq
\label{eq:force}
  \frac{m_p^\star v}{\tpn}\sim \frac{B\,\dB}{4\pi L\, n_e}-\frac{|\dmu|}{L},
\eeq
\beq
   \frac{n_ev}{L}\sim \lambda |\dmu|.
\eeq
These two equations can be solved for $|\dmu|$ and $v$,
\beq
    |\dmu|\sim \frac{B\,\dB}{4\pi n_e(1+L^2/a^2)},
\eeq
\beq
\label{eq:v}
    v\sim \frac{B\,\dB\, \tpn}{4\pi \rho_p L\,(a^2/L^2 + 1)},
\eeq
where $\rho_p=n_em_p^\star\sim \rho/20$ is the mass density of the plasma, and
\beq
   a=\left(\frac{\tpn n_e}{\lambda\,m_p^\star}\right)^{1/2}
\eeq
is a characteristic length introduced by \citet{1992ApJ...395..250G}. 
Its dependence on the electron density $n_e$ is not strong, and we will use
a crude estimate of $n_e$ obtained from the approximate relation
$\tilde{\mu}_e\approx\tilde{\mu}_n\sim 100(\rho/\rhonuc)^{2/3}$~MeV, where
$\tilde{\mu}_e= c\hbar(3\pi^2 n_e)^{1/3}$. This gives
\beq
\label{eq:ne}
   n_e\approx 10^{37}\left(\frac{\rho}{\rhonuc}\right)^2 {\rm ~cm}^{-3},
\eeq
\beq
\label{eq:a}
  a\approx 
  10^4  \,T_9^{-4} \left(\frac{\rho}{\rhonuc}\right)^{5/6} 
        \left(\Rpn\Rlambda H\right)^{-1/2}  {\rm ~cm}.
\eeq
Two regimes are possible: (1) Friction-dominated regime $L\gg a$. The pressure gradient 
is sufficiently quickly erased so that p-n friction is the main factor limiting the drift speed.
(2) Pressure ``pillow'' regime $L\ll a$. Friction is negligible and the magnetic force is 
nearly balanced by the gradient of the local pressure enhancement (``pillow'').
Then the drift speed is controlled by Murca reactions, which tend to deflate the pillow 
and allow slow compression or expansion of the e-p plasma.
 
We end this brief review of ambipolar diffusion with the following remark.
As pointed out by \citet{1992ApJ...395..250G}, solenodial plasma 
motions $\nabla\cdot(n_e\bv)=0$ are not accompanied by any compression
and hence do not perturb $\mu_e$ or $\mu_p$. Such motions
are only limited by the p-n friction,  so in this case the term $a^2/L^2$ in 
\Eq~(\ref{eq:v}) should be removed. The neutron fluid could, in principle, be pulled 
into motion with velocity $\bv_n\neq 0$
without perturbing neutron density or pressure if $\nabla\cdot(n_n\bv_n)=0$.
However, since $n_n(r)\neq n_e(r)$, the incompressible motion could only occur with
$\bv\neq \bv_n$, i.e. neutrons cannot move with the plasma and the p-n friction is 
inevitable. Moreover, the large density of neutrons $n_n\gg n_e$ implies that their 
allowed motions are generally slow compared with those of the plasma, $v_n\ll v$.
Therefore, neutrons are treated as a static background in \Eq~(\ref{eq:fbalance}).

\subsection{Magnetic field evolution equation}\label{sec:eqn}

The magnetic field evolution is governed by the Maxwell equation 
$\partial\bB/\partial t=-c\nabla\times\bE$. The electric field can be expressed from 
the force balance for the electron fluid (omitting the small resistive term),
\beq
\label{eq:el_bal}
  -e\left(\bE+\frac{\bv_e\times\bB}{c}\right)-\frac{\nabla P_e}{n_e}+m_e^\star{\mathbf g}=0. 
\eeq
Here $e$ is the absolute value of the electron charge, $n_e\approx Y_e\rho/m_p^\star$
is the electron density, $\bv_e$ is the velocity of the electron fluid, $P_e\propto n_e^{4/3}$ 
is the electron pressure, ${\mathbf g}=-\nabla \Phi_g$ is the gravitational 
acceleration (in the Newtonian approximation), and $m_e^\star$ is the effective 
inertial mass of the relativistic electron. 
After taking curl of \Eq~(\ref{eq:el_bal}), the two last terms disappear, taking
into account that $(\nabla P_e)/n_e=\nabla(4 P_e/n_e)$ and 
$\nabla m_e^\star\parallel \nabla \Phi_g$. Then one finds
\beq
\label{eq:el1}
   \frac{\partial\bB}{\partial t}=\nabla\times \left(\bv_e\times\bB\right), 
\eeq 
which states that the magnetic field is frozen in the electron fluid.

An alternative form of $\partial\bB/\partial t$ is obtained if $\bE$ is expressed 
from the force balance for the proton fluid,
\beq
\label{eq:ions0}
  e\left(\bE+\frac{\bv\times \bB}{c}\right)+m_p^\star{\mathbf g}-\frac{\nabla P_p}{n_e}
  -\frac{m_p^\star \bv}{\,\tpn}=0,
\eeq
where $P_p$ is the pressure of the degenerate protons, $P_p\propto n_e^{\gamma}$ 
with $\gamma\approx 5/3$, and we have used the neutrality condition ($en_e$ equals 
the proton charge density). This gives
\begin{eqnarray}
\label{eq:ions1}
\nonumber
    \frac{\partial\bB}{\partial t} & = & \nabla\times \left(\bv\times\bB
    -\frac{c\,m_p^\star\bv}{e\,\tpn}\right) \\
    & = &  \nabla\times\left(\bv\times\bB+\bvH\times\bB\right),
\end{eqnarray}
where $\bvH=-\bj/en_e=\bv_e-\bv$ is the Hall velocity (the velocity of the electron fluid
relative to the protons) and in the second equality we have used \Eq~(\ref{eq:fbalance}).
\Eq~(\ref{eq:ions1}) is equivalent to \Eq~(\ref{eq:el1}). 
Note that 
(1) $\bvH$ is perpendicular to the ambipolar drift velocity $\bv$, and 
(2) Hall drift conserves magnetic energy \citep{1992ApJ...395..250G}
while ambipolar drift dissipates it. The ratio of the two drift speeds is given by
\beq
     \frac{v}{\vH}=\frac{\tpn eB}{m_p^\star c}
     \sim 30\, B_{16}T_9^{-2}\left(\frac{\rho}{\rhonuc}\right)^{1/3}\Rpn^{-1}.
\eeq
In the parameter range of main interest $\vH\ll v$.

Note also that the term $\bvH\times\bB$ in \Eq~(\ref{eq:ions1}) is proportional to the 
Lorentz force applied to the e-p plasma, and its solenoidal component 
can only be balanced by the friction force, so Hall drift in the core requires 
p-n friction. The proton fluid itself cannot  offset the force associated with 
Hall drift, because the proton stress tensor $\sigma_{ik}=P_p\delta_{ik}$ is only capable 
of sustaining a curl-free force, which corresponds to a curl-free contribution to the electric 
field and makes no contribution to $\partial\bB/\partial t$.

\subsection{Plateau in the thermal evolution}

A nascent neutron star with its initial temperature $\sim 10^{11}$~K is quickly cooled 
by neutrino emission until heating due to ambipolar drift offsets cooling.
Below we show that the drift of magnetic fields $B\simgt 10^{16}$~G can sustain 
a high temperature $\Tc>6\times 10^8$~K for $\sim 1$~kyr.
We first consider the core with normal (non-superfluid) matter, i.e. assume that 
$\Tcrit$ is below the temperature range of interest.
The core heat capacity is dominated by neutrons and given by 
\beq
   C_V \approx \frac{\pi^2}{2} n_n k\left(\frac{kT}{\tilde{\mu}_n}\right)
          \approx 2\times 10^{20}\,T_9 \left(\frac{\rho}{\rhonuc}\right)^{1/3}
          \frac{\rm erg}{{\rm K~cm}^3},
\eeq
where $k$ is the Boltzmann constant, $n_n\approx \rho/m_n^\star$, and 
$m_n^\star\approx 10^{-24}$~g.

The thermal conductivity of the core is very high, orders of magnitude
higher than in the crust (e.g. \citealp{2001A&A...374..151B,2001MNRAS.324..725G}). 
Any locally generated heat is quickly shared by the entire core at approximately uniform 
temperature $\Tc$.\footnote{In a star with mass $M>M_D$, activation of Durca cooling 
     at the center could create a temperature gradient, however such stars are not 
     considered here --- Durca cooling would steal too much energy and make the 
     core uninteresting as a heat source for the surface luminosity.}
The evolution of $\Tc$ is approximately described by the volume-averaged equation,
\beq
   C_V\frac{d\Tc}{dt}=-\dqnu+\dqh,
\eeq  
where $\dqh$ is the volume-averaged heating rate.
At early times (when $\Tc\gg 10^9$~K) the cooling term strongly dominates, 
$\dqnu\gg\dqh$, and the temperature follows a power-law, 
\beq
\label{eq:cool}
   \Tc\approx 10^9\left(\frac{t}{\rm yr}\right)^{-1/6} {\rm K} \qquad (\dqnu\gg\dqh).
\eeq
The core cools to $10^9$~K in about 1~yr and then the ambipolar drift 
$v\propto \Tc^{-2}$ becomes fast enough to provide strong heating and offset the 
neutrino cooling.

Indeed, consider magnetic field $B$ that varies by $\ddB$ on a scale $L$. 
The scale should not exceed a few km and is certainly smaller than the radius 
of the star; it will be normalized below to $10^5$~cm.
In the temperature range of interest the ambipolar drift occurs in the friction-dominated
regime $a<L$ (see \Eq~(\ref{eq:a})), in contrast to the opposite assumption in TD96
and \citet{2004ApJ...608L..49A}.
The heating rate is the product of the friction force $\rho_p v/\tpn$ and the drift speed $v$,
\beq
   \dqh=\frac{\rho_p v^2}{\tpn}
   \sim \frac{\tpn}{\rho_p}\left(\frac{B\,\ddB}{4\pi L}\right)^2.
\eeq
The heating balances Murca cooling, $\dqh\approx\dqnu$, when the core temperature 
decreases to 
\beq
\label{eq:Tbal}
  \Tbal\approx 
  8 \times 10^8 \left(\frac{B_{16}\,\ddB_{16}}{L_5}\right)^{0.2} 
    \left(\frac{\rho}{\rhonuc}\right)^{-7/30} {\rm~K}.
\eeq
The characteristic timescale for dissipating the available magnetic energy is given by
\begin{eqnarray}
\nonumber
   \tdiss &\sim& \frac{(\delta B)^2}{8\pi \dqh} \sim \frac{2\pi L^2\rho_p}{B^2 \tpn}  \\
     & \approx & 6\times 10^2\,
     \frac{L_5^{1.6} (\ddB_{16})^{0.4}}{B_{16}^{1.6}}  \left(\frac{\rho}{\rhonuc}\right)^{6/5} {\rm ~yr}.
\label{eq:tdiss}
\end{eqnarray}
Comparing with the neutrino cooling timescale $t_\nu=C_VT/2\dqnu$, one finds
\beq
   \frac{\tdiss}{t_\nu} \sim 10\, \frac{L_5^{0.4}(\ddB_{16})^{1.6}}{B_{16}^{0.4}}
      \left(\frac{\rho}{\rhonuc}\right)^{2/15}\gg 1.
\eeq
Thus, strong fields $\delta B\sim B\simgt 10^{16}$~G imply that $\Tc$ stays near 
$\Tbal$ for a relatively long time $\tdiss$, much longer than it takes to reach the balance. 

This picture may be extended to allow a spectrum of magnetic field variations in 
a nascent magnetar,
\beq
  (\ddB)^2\propto L^{-\alpha}, \qquad L<R.
\eeq 
As the core cools and ambipolar diffusion develops, $\ddB$ may be first damped on small 
scales $L$ and then on progressively larger $L(t)$. Equating $\tdiss$ to the stellar age $t$, 
one finds from the above equations $L\propto t^{5/(8-\alpha)}$ and 
$\Tbal\propto L^{-(2+\alpha)/10}$, which gives
\beq
   \Tbal\propto t^{\frac{-(2+\alpha)}{2(8-\alpha)}}.
\eeq 
Eventually ambipolar diffusion becomes efficient on the largest scale $L_{\max}<R$.
Once $\ddB$ is damped on this scale, heating is extinguished and $\Tc$ quickly drops. 
With the end of ambipolar diffusion one may expect a decline in magnetar activity.

\subsection{One-dimensional model}

Ambipolar diffusion may be illustrated by the following model.
Consider an approximately uniform background $\rho\approx const$ and a magnetic 
field in Cartesian coordinates $x,y,z$ of the form,
\beq
\label{eq:B}
   \bB=(0,0,B), \qquad B=B_0\sin \kk x.
\eeq
Ambipolar diffusion will tend to flatten the profile of $B$.
However, the ``null points'' $x=0,\pi/\kk$ where $B=0$ do not move, as 
the magnetic force $\partial/\partial x(B^2/8\pi)$ vanishes at these points. 
As a result, the initial sine profile will relax to the final step-like shape,
\beq
\label{eq:B1}
   B(x)=\pm B_1, \qquad B_1=\frac{2}{\pi}\,B_0,
\eeq
with the jumps at the null points.
Note that a large free energy remains stored in the magnetic field after ambipolar 
diffusion has done its work. As the sine profile relaxes
to the top-hat $B(x)=\pm B_1$, only a fraction $1-8/\pi^2\approx 19$\% of the 
initial magnetic energy is dissipated. The average dissipated
energy density is $U_{\rm diss}\approx 3.8\times 10^{29}\,B_{0,16}^2 {\rm ~erg~cm}^{-3}$.
This value should be compared with the minimum neutrino losses (non-superfluid Murca
cooling) for the desired temperature $T>6\times 10^8$~K over the magnetar age
$t\sim 10^{11}$~s: 
$U_{\rm lost}\simgt 10^{30} (T/6\times 10^8{\rm ~K})^8\, t_{11}$~erg~cm$^{-3}$.
One can see that models with $B_0\simgt 2\times 10^{16}$~G are of main interest for 
the hot-core scenarios. Then the main stage of ambipolar diffusion must occur 
in the friction-dominated regime.

Evolution of the magnetic field is described by \Eq~(\ref{eq:ions1}).
It is easy to see that $\bvH=-(c/4\pi en_e)\nabla\times\bB$ is in the $y$-direction,
$\bvH\times\bB$ is in the $x$-direction, and the Hall term $\nabla\times(\bvH\times\bB)$ 
vanishes. The ambipolar drift velocity of protons (which is along the $x$-axis) leads to
the evolution of $B=B_z$ according to the equation,
\beq
\label{eq:Bdot}
   \frac{\partial B}{\partial t}=-\frac{\partial}{\partial x} (vB). 
\eeq
In the friction-dominated regime, the drift velocity $\bv=(v,0,0)$ is given by
\beq
\label{eq:vfric}
    v= -\frac{\tpn(T)}{\rho_p}\frac{\partial}{\partial x}\frac{B^2}{8\pi}.
\eeq
This yields a nonlinear diffusion equation for $B$.
The diffusion is accompanied by heating with rate 
$\dqh=[\partial_x (B^2/8\pi)]^2 \tpn/\rho_p$, and the temperature evolution is described by
\beq
   C_V\frac{dT}{d t}=\frac{\kk\,\tpn}{2\pi\rho_p}\int_0^{2\pi/\kk}
   \left(\frac{\partial}{\partial x}\frac{B^2}{8\pi}\right)^2 dx - \dqnu.
\label{eq:T}   
\eeq
The coupled \Eqs~(\ref{eq:B})-(\ref{eq:T}) can be solved numerically for 
$B(t,x)$and $T(t)$. Note, however that these equations assume $\dmu\approx 0$
due to efficient Murca reactions and do not take into account the possible 
build-up of a pressure gradient (pillow) as the flow converges toward the null points.

An approximate solution to the full problem, which includes the pillow 
formation, may be obtained as follows. Let us define a characteristic scale 
\beq
  \Leff=B_1\left(\frac{dB}{dx}\right)^{-1},
\eeq
with $dB/dx$ evaluated at the null point $x=0$.
In the region $0<x<\Leff$ we have $B(x)\approx B_1 x/\Leff$. 
The initial $B(x)=B_0\sin\kk x$ has $L_1=2/\pi \kk$, and later $L_1$ shrinks --- 
the profile $B(x)$ steepens near the null point as 
it evolves toward the final top-hat shape $B(x)=\pm B_1$.
Using magnetic flux conservation, one can parameterize the state of the system at 
any time $t$ with only one degree of freedom $L_1(t)$ (see Appendix), which obeys 
the following dynamic equation,
\begin{eqnarray}
\label{eq:Ldot}
  \frac{dL_1}{dt}\approx \left\{\begin{array}{ll}
\vspace*{0.2cm}
      \displaystyle{-\frac{\tpn B_1^2}{2\pi\rho_p L_1}} & \quad L_1>\Ls, \\
      \displaystyle{-\frac{\lambda B_1^2 L_1}{4\pi n_e^2}} & \quad L_1<\Ls.
                           \end{array}
                   \right.
\end{eqnarray}
Here the transition $L_1=\Ls$ corresponds to $L_1/a=\sqrt{2}$. At this moment,
the rate of Murca reactions becomes insufficient to remove $\dmu$ in the compressed
region near $x=0$, and the dynamics near the null point occurs in the pillow-dominated 
rather than friction-dominated regime.
The coefficient $\lambda$ is evaluated inside the pillow at $x=0$, where the local 
$\dmu$ can exceed $kT$. Therefore, $\lambda$ in \Eq~(\ref{eq:Ldot}) must be 
calculated using the correction factor $H(\xi)$ (see \Eq~(\ref{eq:lambda}) and Appendix~A).

The evolution of the system is described by two coupled differential equations for 
$L_1(t)$ and $T(t)$. The temperature $T(t)$ remains approximately uniform across 
the domain, because of the high thermal conductivity, and its evolution is described by 
\beq
\label{eq:Tdot}
    C_V\, \frac{dT}{dt}=\dqh-\dqnu, \qquad  \dqh \approx -\frac{B_1^2\, \kk \dot{L}_1}{12\pi^2}.
\eeq 
This approximate expression for the volume-averaged 
heating rate $\dqh$ is derived in Appendix~A. It underestimates the heating rate by 
a factor of 2 at the initial stage when $L_1=2/\pi\kk$. A simple approximate way to 
correct this (used in the numerical models below) is to multiply $\dqh$ by $1+\pi\kk L_1/2$.
The volume-average cooling rate $\dqnu$ is dominated by the large region 
$x>L_1$ where $\Delta\mu<kT$ 
throughout the evolution; therefore, the standard Murca cooling can be used for $\dqnu$
(\Eq~\ref{eq:Murca}), neglecting $\dmu$.

\begin{figure}[t]
\vspace*{-0.3cm}
\begin{tabular}{c}
\hspace*{-1.5cm}
\includegraphics[width=0.65\textwidth]{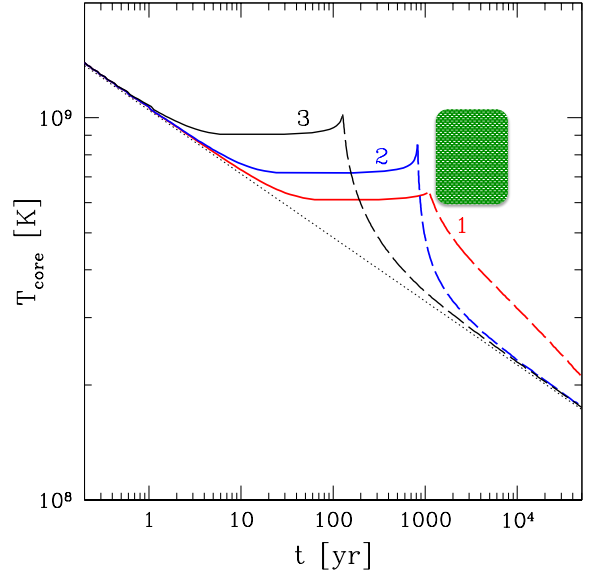} 
\end{tabular}
\vspace*{-0.8cm}
\caption{Temperature evolution in a non-superfluid core with the simple initial 
configuration of the magnetic field given in \Eq~(\ref{eq:B}). The curve $\Tc(t)$ 
is solid as long as ambipolar drift proceeds in the friction-dominated regime in 
the entire domain, and dashed after the formation of pressure pillows at the 
null points. Three sample models are shown:
(1) $B_1=10^{16}$~G  ($B_0=1.57\times 10^{16}$~G), 
$\kk=\pi\times 10^{-6}$~cm$^{-1}$ (red curve),
(2) $B_1=1.5\times 10^{16}$~G  ($B_0=2.36\times 10^{16}$~G),
$\kk=\pi\times 10^{-6}$~cm$^{-1}$ (blue curve), and
(3) $B_1=1.5\times 10^{16}$~G, $\kk=10^{-5}$~cm$^{-1}$ (black curve).
All models assume $\rho=\rhonuc$.
The dotted line shows the core cooling in the absence of heating by ambipolar 
diffusion (\Eq~\ref{eq:cool}). 
The region that would explain the observed surface luminosities 
$\Lum_s\approx 10^{35}$~erg~s$^{-1}$ at ages 1-10~kyr is shaded in green.
 }
\label{fig:Tc}
\end{figure}

Figure~\ref{fig:Tc}  shows the temperature evolution calculated in a few sample 
models with $B_1=10^{16}$~G and $1.5\times 10^{16}$~G.
One can see that the cooling curve $T\approx 10^9\,t_{\rm yr}^{-1/6}$~K 
is followed by the heating=cooling plateau with $T=\Tbal$.
The models with $\Tbal>6\times 10^8$~K have the plateau duration up to $\sim 1$~kyr.
The higher the plateau temperature $\Tbal$ the shorter its duration.
We ran many more models with various $B_0$, $\rho$, and $2\pi/\kk\leq 20$~km,
and in all cases the core temperature was below $6\times 10^8$~K at the 
typical observed magnetar age $t=1-10$~kyr. 
This tension between the model and observations would only be alleviated for 
smaller $\kk$ outside the plausible range $2\pi/\kk\leq 20$~km.

The heating stage is followed by a steep drop of temperature back to the cooling curve 
$T\approx 10^9\,t_{\rm yr}^{-1/6}$~K.
Note that the temperature evolution shown in Figure~\ref{fig:Tc} 
does not take into account the transition to neutron superfluidity, which should
occur when the temperature drops well below $10^9$~K. This transition is accompanied 
by enhanced cooling due to Cooper pair formation 
\citep{2009ASSL..357..247P}, and the core 
temperature will decrease to $10^8$~K much faster than in million years.

The simple one-dimensional model illustrates another interesting feature of 
ambipolar diffusion: the creation of current sheets separating the domains of opposite 
magnetic fields. It is described in more detail in Appendix~A.
The appearance of current sheets may be viewed as a consequence of magnetic 
flux conservation: ambipolar diffusion tends to minimize the magnetic energy 
while the magnetic flux remains frozen in the plasma. Current sheets are also expected 
in MHD relaxation of more general (less symmetric) magnetic configurations 
(e.g. \citealp{2009arXiv0909.1815G,2015MNRAS.450.3201B}) and can have a strong 
guide field.

\subsection{Effects of superconductivity and superfluidity}

The models in Figure~\ref{fig:Tc} neglect possible superconductivity near the 
null point, where the magnetic field is weak and incapable of suppressing 
Cooper pairing of protons. The superconducting region, where $B<B_c$,
has the thickness $L_c\approx (B_c/B_1)L_1$. 
Here the magnetic flux becomes quantized into flux tubes, which 
reduces the effective magnetic pressure. On the other hand, 
superconductivity also suppresses Murca reactions and so
the region $x<L_c$ becomes nearly incompressible.
This will prevent the collapse of the current sheet, 
however will not change our conclusions regarding the temperature evolution.
Superconductivity near null points reduces the energy dissipated by ambipolar 
diffusion and does not help to achieve $6\times 10^8$~K at ages of 1-10~kyr. 
Note also that in a less symmetric configuration, with a guide magnetic field
in the current sheet, $B$ would not go through zero and can be strong 
enough to quench superconductivity everywhere.

Next consider the effects of neutron superfluidity.
The critical temperature for Copper pairing of neutrons is lower than that for protons, but may be high enough to interfere the evolution at temperatures 
$T\simlt 10^9$~K. Neutron superfluidity brings the following changes:

\noindent
(1) The rate of p-n collisions is reduced by the factor $\Rpn< 1$.
This reduction promotes ambipolar diffusion.

\noindent
(2) Superfluidity suppresses Murca reactions
responsible for erasing $\dmu$ by the factor $\Rlambda<1$; this slows down 
the compressive ambipolar drift.

\noindent
(3) Although the Murca cooling is suppressed,
a much stronger cooling occurs due to 
Cooper pairing at temperatures $0.3<T/\Tcrit<1$.
It implies a cooling phase with $\dot{q}_\nu$ exceeding 
$10^{22}T\,_9^8$~erg~s$^{-1}$~cm$^{-3}$ at $T\approx (0.7-0.8)\Tcrit$ 
(see Figure~\ref{fig_qv} and \citealp{2009ApJ...707.1131P}). 

\noindent
(4) Superfluid neutrons lose their heat capacity. The heat capacity of the core 
can become dominated by protons, which
are guarded from Cooper pairing by the ultra-strong field $B\simgt 10^{16}$~G. 

As soon as $T$ decreases below $\Tcrit$ the strong Cooper pair cooling switches 
on and the heating cannot balance it until $T/\Tcrit\sim 0.3-0.5$.
At these temperatures, the suppression factors $\Rpn$ and $\Rlambda$ are
moderate --- both are comparable to 0.2. 
Most of the dissipation still occurs in the friction-dominated regime,\footnote{Superfluidity 
         increases $a$ by the factor $(\Rpn\Rlambda)^{-1/2}\sim 5$ during the p-n 
         friction stage. On the other hand, the faster heating implies a higher temperature,
         which tends to reduce $a$ as $a\propto T^{-4}$ (\Eq~\ref{eq:a}). 
         }
and the main effect of superfluidity is the increased dissipation rate,
shortening the duration of the main heating by the factor of $\sim 0.2-0.3$.
Superfluidity only makes the final (pillow) stage
slower, as it reduces $\lambda$ and makes the pillow harder,
however the heating at this stage is insufficient to sustain $\Tc>6\times 10^8$~K.
Therefore, superfluidity does not help the core to become the main heat source
for persistent magnetars.

\subsection{Comparison with previous work}

In contrast to \citetalias{1996ApJ...473..322T}, we find that the plateau phase 
(the balance between ambipolar heating and neutrino cooling) does not sustain 
$\Lum_s\approx 10^{35}$~erg~s$^{-1}$ for 10~kyr. 
The main reason for this disagreement is the heating mechanism. 
\citetalias{1996ApJ...473..322T} assumed that ambipolar drift occurs in the pillow regime, 
i.e. it is limited by the finite rate of Murca reactions, sustaining the pressure pillow 
$\dmu\sim B^2/8\pi$. In this case, what \citetalias{1996ApJ...473..322T} call heating 
and cooling processes are in fact the same Murca process that converts 
$e,p\leftrightarrow n$ while changing temperature and producing
neutrinos.\footnote{To clarify the meaning of the thermal balance in the pillow regime 
       one should note the following. 
       Murca reactions are pure cooling when $\dmu\ll kT$ and pure heating when 
       $\dmu\gg kT$. The latter limit is approached when $\dmu>10 kT$ --- then each 
       Murca reaction releases energy $\dmu$; 3/8 of this energy is carried away by 
       neutrinos and 5/8 heats the matter \citep{2006MNRAS.372..276F}. TD96 simply
       assumed that in thermal balance $kT\approx \dmu$. However, in the pillow regime 
       of ambipolar diffusion, there is a strong gradient of $\dmu$ while $T$ is approximately 
       uniform due to efficient heat conduction. In this situation, heating=cooling means the balance 
       between  Murca heating in the regions of large $\dmu/kT$ and Murca cooling in the 
       regions of small $\dmu/kT$. As the field evolves, heating tends to concentrate in a 
       small fraction of the core volume (see the end of \Sect~3.4).}
In contrast, we find that ambipolar diffusion could sustain 
$\Lum_s\approx 10^{35}$~erg~s$^{-1}$ only when it occurs in the friction-dominated 
regime, i.e. when $\dmu$ is unimportant. The heating by p-n friction is capable of offsetting
the neutrino cooling at $\Tc>6\times 10^8$~K, however this balance has a short lifetime.
The suggestion of \citetalias{1996ApJ...473..322T} that neutron superfluidity would 
prolong the hot phase is incorrect; they neglected the Cooper pair cooling.

\citet{2004ApJ...608L..49A} extended the model of  \citetalias{1996ApJ...473..322T}
by assuming superconductivity at $T<\Tcrit=5\times 10^9$~K and by including Hall drift 
(we find that Hall drift is unimportant in the core, see \Sect~\ref{sec:eqn}).
Superconductivity would suppress the Murca reaction by a factor of $\sim 10^2$
before the core temperature drops to $T\sim 7\times 10^8$~K \citep{2001A&A...372..130H}.
Then it becomes possible to sustain this temperature for a long time, 
because neutrino cooling becomes slow:
cooling due to Murca and Cooper pairing of protons are both inefficient at $T\ll \Tcrit$,
and cooling due to Cooper pairing of neutrons may not begin yet at $T\sim 7\times 10^8$~K.
Superconductivity {\it everywhere} in the core is an essential assumption of this picture.
We argued, however, that the energy 
budget of magnetars implies that $B\sim 10^{16}$~G somewhere inside the star, 
quenching superconductivity. Then neutrino cooling cannot be suppressed at
$\Tc\sim 10^9$~K. Note that quenching superconductivity in a fraction of the core 
volume is sufficient for fast cooling of the entire core. Quenching is particularly easy 
in the outer core, as this requires field $B_c<10^{16}$~G.

\citet{2011MNRAS.413.2021G} studied in detail the effect of strong superfluidity 
on ambipolar diffusion. They focused on the regime $\Tc\ll\Tcrit$, which permits 
simple analytical expressions for the suppression factors $\Rpn$ and $\Rlambda$. 
This asymptotic description is useful for superfluid 
neutrons in a cool core (with normal protons). 
However, it is not applicable to the main phase of ambipolar 
diffusion that releases most of the energy --- in the temperature range of main interest, 
$T>6\times 10^8$~K, $\Tcrit/T$ can hardly exceed 3.

\citet{2012MNRAS.422.2632H} calculated the temperature of a core heated 
by the decay of an initial $B=10^{16}$~G on a prescribed timescale of 10~kyr.
This phenomenological heating model gave $\Tc\approx 7\times 10^8$~K at 1~kyr 
and $5\times 10^8$~K at 10~kyr. 
They deemed $\Tc\approx 7\times 10^8$~K insufficient because it gave $T_s$
below the spectroscopically measured surface temperature $T_X$
(after correcting for the gravitational redshift). In fact, $T_s=(F_s/\sigSB)^{1/4}$ is 
allowed to be somewhat below $T_X$ as the surface emission 
deviates from blackbody due to radiative transfer effects in the surface layers.


\section{Thermal balance for a heated crust}
\label{balance}

We now turn to another possible explanation of the high surface temperature: 
a dissipative process in the crust of the neutron star. General requirements to a 
successful quasi-steady heater in the crust were investigated by 
\citet{2006MNRAS.371..477K,2009MNRAS.395.2257K,2014MNRAS.442.3484K}. 
They assumed a cool core and placed a phenomenological heat source at various 
depths in the crust without specifying its mechanism. Their detailed simulations of 
heat conduction and neutrino cooling demonstrated that a heating rate
$\dqh\simgt 3\times 10^{19}$~erg~s$^{-1}$~cm$^{-3}$ is required at depths 
$z<300$~m to sustain the surface luminosity $\Lum_s\approx 10^{35}$~erg~s$^{-1}$. 

Our goal is to assess if physical mechanisms --- mechanical or ohmic dissipation --- 
could provide such heating. However, we begin with a simple phenomenological model 
similar to that of \citet{2014MNRAS.442.3484K} to check the constraints on the required 
heating. Our sample numerical models below assume a neutron star 
with a canonical mass $M=1.4M_\odot$ and the BSk20 equation of state $P(\rho)$ 
\citep{2013A&A...560A..48P}; it has the radius $R=11.7$~km and 
surface gravity $g=1.7\times 10^{14}$~cm~s$^{-2}$.

In the presence of steady crustal heating, the heat transfer equation reads,
\beq
\label{eq:steady}
   -\frac{d}{dz}\left( \kappa\frac{dT}{dz}\right) = \dot{q}_h-\dot{q}_\nu.
\eeq
It determines the subsurface temperature profile $T(z)$ for a given 
heating rate $\dot{q}_h(z)$ and the self-consistently calculated neutrino cooling 
rate $\dot{q}_\nu(z,T(z))$.
The crust is approximated as a slab of thickness much smaller than the stellar
radius; then the relativistic metric coefficients may be approximated as constant 
and cancelled from the heat transfer equation.
We numerically solve \Eq~(\ref{eq:steady}) as described in \citet{2015ApJ...815...25L},
using thermal conductivity $\kappa$ calculated by Potekhin's code
\citep{1999A&A...351..787P} and neutrino emissivities given by \citet{2001PhR...354....1Y}.
The solution with $\dqh=0$ gives the relation between $T_s$ and temperature 
$T_b$ at a chosen depth $z_b$ above the heater. We choose a small
$z_b\approx 60$~m where $\rho_b=10^9$~g~cm$^{-3}$ and use the obtained 
$T_b$-$T_s$ relation as a boundary condition in models with heating at $z>z_b$.

Below we examine the ability of crustal heating to power the observed $\Lum_s$ 
and therefore consider models with a relatively cool core $\Tc\ll 10^9$~K which is not 
capable of sustaining $\Lum_s=10^{35}$~erg~s$^{-1}$.
In the sample models we assume $\Tc=2\times 10^8$~K,
which sustains a surface luminosity $\Lum_s\sim 3\times 10^{33}$~erg~s$^{-1}$ with 
an iron envelope and $\Lum_s\sim 10^{34}$~erg~s$^{-1}$ with a light-element envelope 
(Figure~\ref{fig_Tc}).

\begin{figure}[t]
\vspace*{-2cm}
\begin{tabular}{c}
\includegraphics[width=0.45\textwidth]{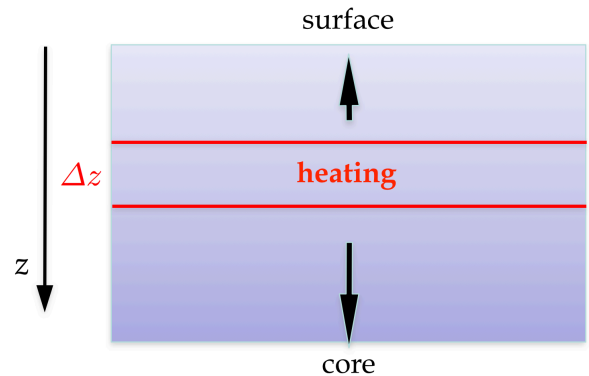} 
\end{tabular}
\vspace*{-2.5cm}
\caption{A heated layer of thickness $\Delta z$ at depth $z_h$ feeds the heat 
flux toward the core $F_{\rm down}$ and toward the stellar surface $F_{\rm up}$. 
The net heating rate per unit area is $F_{\rm up}+F_{\rm down}=F_h=\dqh \Delta z$. 
 }
 \label{fig:source}
 \end{figure}

The desired surface flux $F_s=\sigSB T_s^4$ requires a heating rate per unit area,
\beq
   \Fh=\int \dqh~dz\gg\Fs,
\eeq
as most of the heat is conducted to the core and lost to neutrino 
emission; only a small fraction $\eff$ is conducted to the surface
(Figure~\ref{fig:source}). The required $\Fh$ depends on 
the characteristic depth $z_h$ where heating occurs. 
The calculation is simplified if we use the approximation of a thin 
heated layer with thickness $\Delta z\ll z_h$,
\beq
\label{eq:delta}
   \dot{q}_h=\Fh\, \delta(z-z_h).
\eeq
This idealized model gives a reasonable approximation to the required $F_h$, 
which is independent of $\Delta z$. The value of $\Delta z\simlt z_h$ is used
to convert the results obtained with the delta-function approximation to a realistic 
heating rate, using the relation $\dqh=F_h/\Delta z$.

The solution of \Eq~(\ref{eq:steady}) with the heat source (\ref{eq:delta})
is found as follows. We fix the effective surface temperature $T_s=4\times 10^6$~K 
(which corresponds to $\Fs=\sigSB T_s^4\approx 10^{22}$~erg~s$^{-1}$~cm$^{-2}$)
and integrate the heat diffusion equation 
with $\dqh=0$ downward to $z_h$ where the heater is located. Thus we find $T(z_h)$ 
and the heat flux $F_{\rm up}$ from $z_h$. This flux can be somewhat larger than $\Fs$, 
because of neutrino losses at $z<z_h$. The heating rate $F_h$ at $z_h$ feeds two fluxes:
toward the surface and toward the core, $\Fh=F_{\rm up}+F_{\rm down}$. 
We find the downward flux $F_{\rm down}(z_h)$ using iterations:
any trial $F_{\rm down}$ gives a steady solution connecting $T(z_h)$ and $\Tc$,
and we iterate it until the solution matches $\Tc=2\times 10^8$~K
at the bottom of the crust, $\rho=1.4\times10^{14}$~g~cm$^{-3}$.
Note that 
$\dqnu\neq 0$ and part of the heat flux is lost to neutrino emission before 
reaching the core. 

\begin{figure}[t]
\begin{tabular}{c}
\hspace*{-0.7cm}
\includegraphics[width=0.55\textwidth]{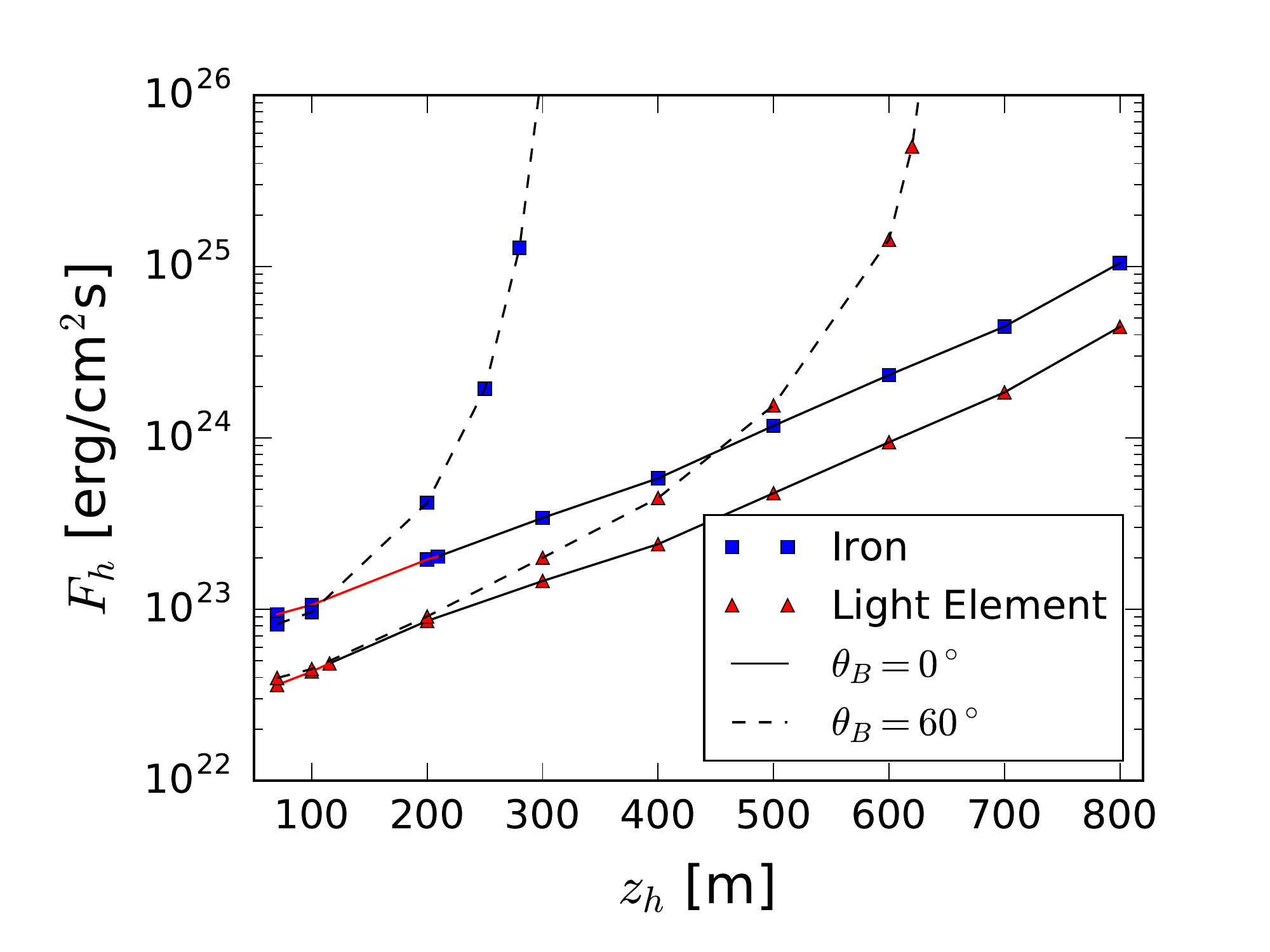} 
\end{tabular}
\vspace*{-3mm}
\caption{
The required internal heating rate per unit area of the crust, $\Fh$, as a function 
of the heater position $z_h$.
Each symbol shows a calculated model of steady heat transfer from the heater 
to the stellar surface (and to the core of temperature $\Tc=2\times 10^8$~K).
In all models, $\Fh$ is adjusted to sustain the effective surface temperature 
$T_s=4\times 10^6$~K, which corresponds to surface flux 
$F_s\approx 10^{22}$~erg~s$^{-1}$~cm$^{-2}$. Magnetic field $B=10^{15}$~G is 
assumed and two cases are shown: $\thB=0$ (radial field) and 
$\thB=60^{\rm o}$. The calculations are performed for two different chemical 
compositions of the envelope --- iron and maximal light element envelope.
The melted region is indicated by the red part of the curve connecting the symbols.
 }
 \label{fig:Fh}
 \end{figure}

The result for $F_h=F_{\rm up}+F_{\rm down}$ is shown in Figure~\ref{fig:Fh}. 
One can see that $\Fh\gg F_s\approx 10^{22}$~erg~s$^{-1}$~cm$^{-2}$ is 
required in all cases except when the heater is very close to the surface
(near or outside the boundary of our computational domain $z_b\approx 60$~m.)
A moderate inclination of the magnetic field significantly reduces the radial heat flow.
Inclination $\thB=60^{\rm o}$ strongly increases the required $F_h$, especially
for the iron envelope, and excludes $z_h\gg 100$~m. The steep increase and runaway 
of the required $\Fh$ at large $z_h$ is the result of neutrino losses, which 
prevent the internal temperature profile $T(z)$ from reaching the values required 
to sustain $F_s$.


\section{Mechanical heating}

The ultrastrong magnetic fields of magnetars can stress their crusts beyond the
elastic limit \citepalias{1996ApJ...473..322T}. Then the crustal deformations become 
irreversible and are accompanied by heating. 
Part of the released magnetic energy is passed to the external 
magnetosphere attached to the crust and part is converted locally to heat. 
Thermoplastic waves  effectively ``burn'' magnetic energy in the crust, 
resembling deflagration fronts in combustion.

Large stresses can be created in the crust in three ways:
\\
(1) Magnetic field evolution in the liquid core differs from the field behavior in the solid 
crust. This generates a gradient in the field at the crust-core interface. The resulting 
magnetic force applied to the crust may be able to deform it beyond the elastic limit. 
Then the crust is expected to experience a shear flow, relieving the applied stress.
This shear flow will tend to localize along ``heat lines'' similar to those observed in 
laboratory experiments with a torsional Hopkinson bar (e.g. \citealp{2002pmas.book.....W}). 
It must, however, 
satisfy an important constraint: the crustal shear should not tear magnetic field lines 
(as this would generate magnetic energy) and may develop along magnetic flux surfaces.
\\
(2) Magnetic stresses can be fostered by internal processes in the crust itself,
in particular due to Hall drift.
As long as ohmic dissipation is negligible, the magnetic field remains frozen in the 
electron fluid drifting through the ion lattice with velocity $\bv_{\rm H}=\bj/en_e$,
where $\bj=(c/4\pi)\nabla\times\bB$ is the electric current density determined 
by the magnetic configuration of the star.
The Hall drift $\bvH$ deforms the magnetic field lines and is capable of creating 
large magnetic stresses 
(\citetalias{1996ApJ...473..322T}; \citealp{2011ApJ...741..123P}).
This leads to launching thermoplastic waves \citep{2014ApJ...794L..24B}, 
which move the crust and relieve the internal magnetic stresses. 
\citet{2016arXiv160604895L} further investigate plastic flows fostered by Hall drift and find 
that they can occur in avalanches that develop due to the
excitation of short Hall waves by the plastic flows.
\\
(3) 
Magnetospheric flares launch strong Alfv\'en waves that are ducted 
along the magnetic field lines and impinge on the crust.
The waves carry enormous magnetic stresses that immediately initiate a 
strong oscillating plastic flow in the crust until the wave is 
damped into heat, which occurs on a timescale of $\sim 10$~ms \citep{2015ApJ...815...25L}.  

Below we explore the maximum efficiency of magnetar surface heating by 
mechanical dissipation in the crust. 
It must satisfy two general constraints:

\noindent
(1) Mechanical dissipation can only occur in the solid phase below the ocean.
At shallow depths $z\simlt 100$~m the crust is melted and forms a 
liquid ocean with a negligible shear viscosity.
This fact limits the efficiency of heating the surface, because most of the heat 
produced at large depths is conducted to the core and lost to neutrino emission.

\noindent
(2) The mechanical heating rate is proportional to the shear stress of the deformed 
crust. There is an upper limit on this stress (maximal strength of the crustal lattice) 
which imposes a ceiling on the heating rate.

\subsection{Quasi-steady mechanical heating}\label{5.1}

We first examine whether quasi-steady mechanical dissipation can explain the 
surface luminosity of persistent magnetars, $\Lum_s\approx 10^{35}$~erg~s$^{-1}$.

Since there is no mechanical heating in the ocean, one can find
its temperature profile from \Eq~(\ref{eq:steady}) with $\dqh=0$ (for a given 
$T_s$). This profile determines the melting depth $\zm$ --- the bottom of the ocean 
--- where $T$ reaches the melting temperature 
$\Tm(\rho)\approx 2.4\times 10^9\,\rho_{12}^{1/3}$~K.  
For instance, for an iron envelope with a radial magnetic field, the surface 
temperature $T_s=4\times 10^6$~K implies $\zm\approx 200$~m.
The ocean is less deep, $\zm<60$~m, for the light-element envelope.

A conservative lower limit on $F_h$ required to sustain 
$T_s\approx 4\times 10^6$~K 
is obtained by assuming that mechanical heating is 
concentrated at the shallowest possible depth, i.e. $\dqh$ is given by
\Eq~(\ref{eq:delta}) with $z_h=\zm$.
A realistic $\dot{q}_h$ must be distributed over a range 
of depths $z>\zm$ (and $\dqh$ is bounded from above, as discussed below), so realistic 
mechanical heating will be less efficient in feeding the surface flux $F_s$. 
Therefore, the model with $\dqh=F_h\,\delta(z-\zm)$
gives a conservative upper limit on the surface heating
efficiency $\eff=F_s/\Fh$.

\begin{figure}[t]
\hspace*{-6mm}
\begin{tabular}{c}
\includegraphics[width=0.55\textwidth]{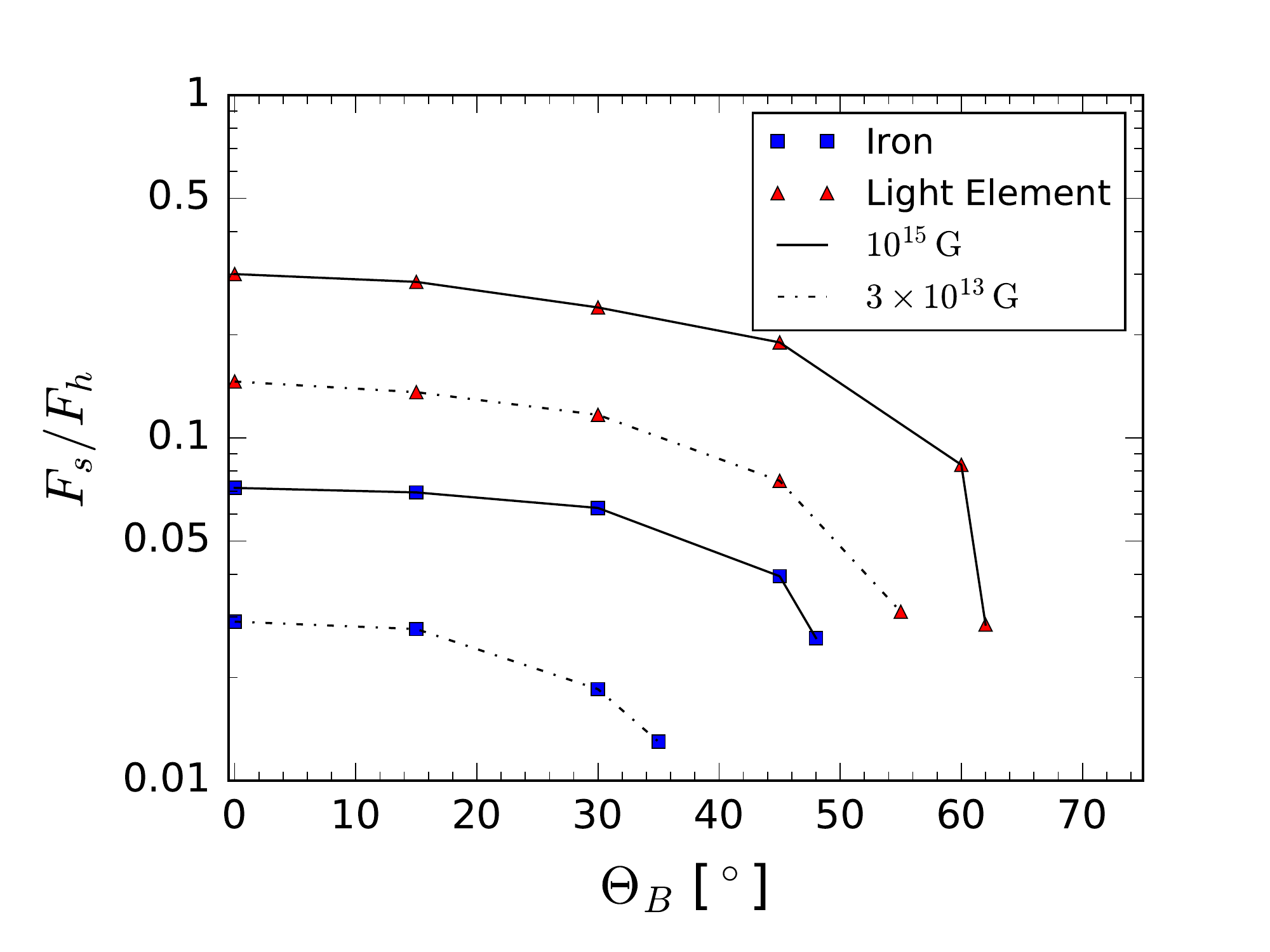} 
\end{tabular}
\caption{Fraction of heat conducted to the stellar surface,
$\eff=\Fs/\Fh$, from a steady delta-function heat source located at $\zm$.
All calculated models have $\Fs=10^{22}$~erg~s$^{-1}$~cm$^{-2}$; the star is 
assumed to have a relatively cool core ($\Tc=2\times 10^8$~K).
The surface heating efficiency $\eff$ is shown as a function of magnetic field angle 
$\thB=\arccos (B_r/B)$. The main fraction $1-\eff$ of the produced heat is conducted 
to the core and lost to neutrino emission.
}
 \label{fig:eff}
 \end{figure}

The efficiency of the heater localized at $\zm$
is shown in Figure~\ref{fig:eff} for $\Fs\approx 10^{22}$~erg~s$^{-1}$~cm$^{-2}$.
Replacing the delta-function with a more realistic heating $\dqh$ distributed over 
$z>\zm$ would significantly reduce $\eff$, and a tilt of magnetic field $\thB>0$ would 
further reduce $\eff$. We conclude that the most optimistic $\eff\simlt 0.1$.

It is also useful to estimate the energy budget invoked by the crustal heating scenario.
Using the typical age of  magnetars, $t\sim 10^{11}$~s, their emitted energy from the 
surface is $E_s\sim \Lum_s t\sim 10^{46}$~erg. The modest efficiency of surface 
heating implies deposition of significant energy in the crust,
\beq
\label{eq:Eh}
   E_h\sim 10^{48}\left(\frac{\epsilon}{0.01}\right)^{-1} {\rm~erg}.
\eeq

Our next goal is to compare the required heating rate with the maximum rate of 
mechanical dissipation due to crustal motions. Note that vertical motions are arrested
by the hydrostatic balance between two dominant forces --- gravity and pressure gradient. 
The pressure $P$ of the compressed, hydrostatic crust is dominated by 
degenerate electrons (or neutrons, near the bottom of the crust).
The lattice Coulomb energy density $U_{\rm Coul}$ is much smaller than $P$
and the crust is relatively fragile to horizontal shear, which leaves pressure unperturbed.
Therefore, we consider below dissipation due to horizontal shear motions.

The dissipative flow of the lattice begins 
when its elastic shear stress $\sigma$ reaches 
a critical value $\sigcr$.
The highest possible value of $\sigcr\sim 0.1\mu$ represents the strength of an ideal 
crystal subject to a fast shear deformation, 
where $\mu$ is the shear modulus of the lattice.
The flow initiated in response to excessive stress buffers stress growth and satisfies
the condition
\beq
\label{eq:sigmax}
  \sigma<\sigma_{\max}\sim 0.1\mu.
\eeq
This is a conservative limit, which may only be approached when the crust is cold and 
deformed quickly \citep{2010MNRAS.407L..54C}. 
Note that $\mu$ is comparable to $U_{\rm Coul}$ and 
the maximum lattice stress is always a fraction of $\mu$, because
there is no agent to carry the stress other than the Coulomb fields.

The rate of mechanical dissipation is given by
\beq 
  \dot{q}_h=-\sigma\dspl, 
\eeq 
where $\spl$ is the strain of the dissipative (plastic) deformation, and $\dspl$ is its 
time derivative. The time-averaged $\dspl$ driven by magnetic field evolution
inside the star may be estimated as follows. The solid crust serves as a gate 
for the energy strored in helical magnetic fields inside the star 
\citepalias{1996ApJ...473..322T}.
The stored wound-up field $B$ can significantly 
exceed its radial component $B_r$ emerging through the stellar surface, possibly by a 
factor up to $\sim 10^2$ (which corresponds to $B\simlt 10^2B_r\sim 10^{16}-10^{17}$~G).
The maximum angle of field unwinding 
$B_{\max}/B_r\sim 10^2$~radian gives a net maximum strain flow $s_{\max}\sim 10^2$.
The corresponding maximum {\it average} strain rate over the active lifetime of a 
magnetar $t\sim 10^{11}$~s is 
\beq
\label{eq:smax}
  \dsav\sim \frac{s_{\max}}{t}\sim 10^{-9} {\rm ~sec}^{-1}\approx 0.03 {\rm~yr}^{-1}.
\eeq

The maximum stress of a plastic flow $\sigma_{\max}\sim 0.1\mu$ gives the maximum 
heating rate,
\beq
\label{eq:dqmax}
   \dot{q}_h^{\max}=\sigma_{\max} \,\dot{s}\sim 10^{18}\, \rho_{12}\, 
   \dot{s}_{-9}
       {\rm ~erg~s}^{-1}{\rm cm}^{-3}.
\eeq
Here we used $\mu\approx 10^{28}\rho_{12}$~erg~cm$^{-3}$ 
(e.g. \citealp{1991ApJ...375..679S}); in the numerical models below we use 
more detailed approximations for $\mu$ from
\citet{2005ApJ...634L.153P} and \citet{2007MNRAS.375..261S}.

The characteristic scale of density variation with depth is $\Delta z\approx 100$~m 
for depths $z$ of interest, including the lower crust.
The heating in a layer around a given $\rho$ may be estimated as 
$F_h(\rho)=\dqh(\rho)\,\Delta z$. This gives the energy release rate per unit area of the crust,
\beq
\label{eq:Fpl}
   F_h 
   \sim 10^{22}\,\rho_{12}\,
                 \left(\frac{\sigma}{\sigma_{\max}}\right)
                  \dot{s}_{-9}
      {\rm ~erg~s}^{-1}{\rm cm}^{-2}.
\eeq
This shows 
that even with the most optimistic assumptions, quasi-steady mechanical 
dissipation can hardly provide the needed heat source 
$F_h\sim 10^{24}(\epsilon/0.01)^{-1}$~erg~s$^{-1}$~cm$^{-2}$ capable of 
powering the observed surface luminosity. The upper bound on $F_h$ is somewhat lifted
to $\sim10^{24}$~erg~s$^{-1}$~cm$^{-2}$ if the plastic flow occurs in the deep crust
where $\rho\sim 10^{14}$~g~cm$^{-3}$. However, this remains insufficient as the 
efficiency of surface heating by the deep heat source decreases below $10^{-2}$
(cf. Figure~\ref{fig:Fh}).
We conclude that quasi-steady mechanical dissipation is incapable of powering 
the persistent surface luminosity of bright magnetars.  

\begin{figure}[t]
\hspace*{-3mm}
\begin{tabular}{c}
\includegraphics[width=0.5\textwidth]{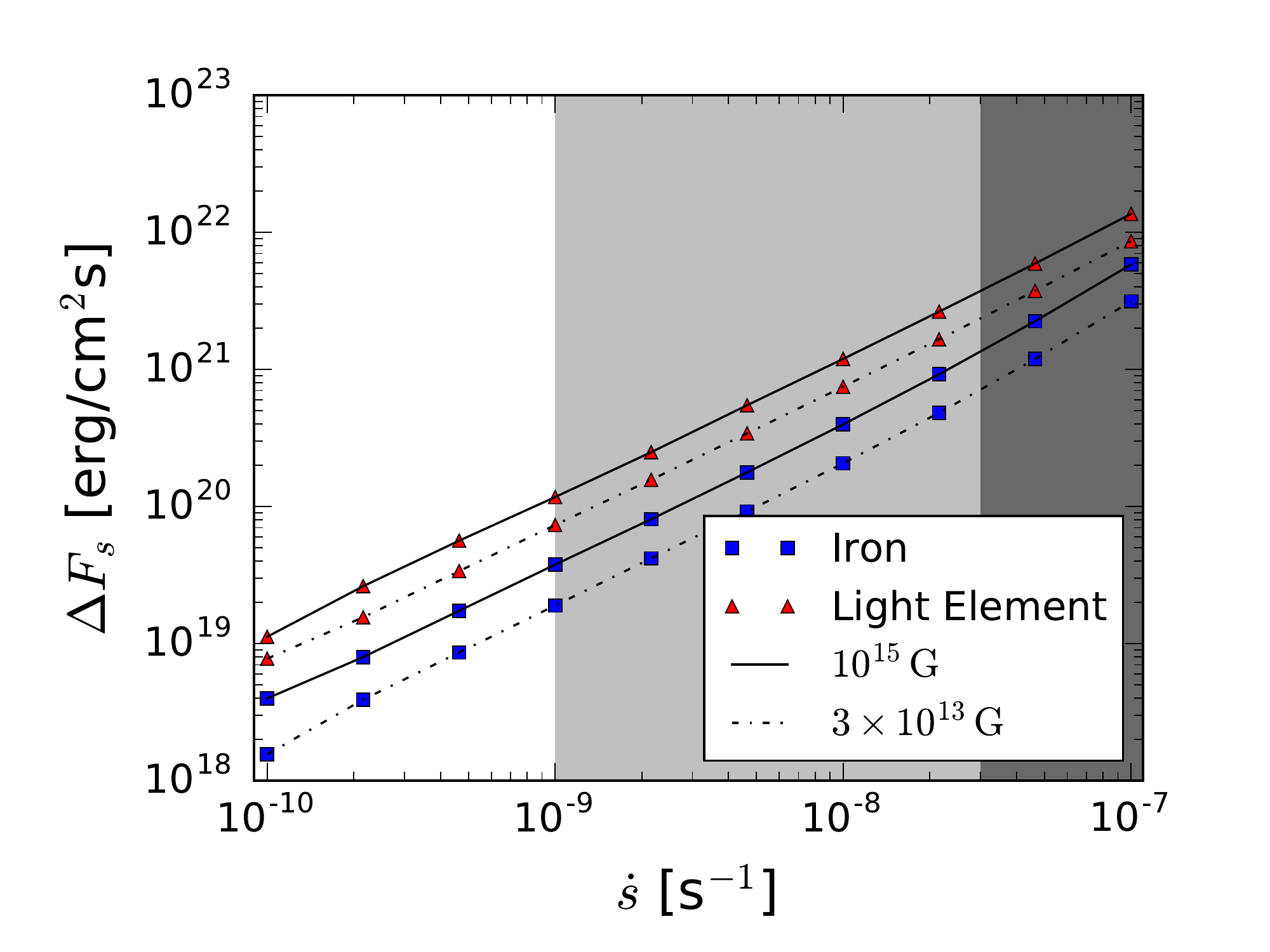} \\ 
\includegraphics[width=0.5\textwidth]{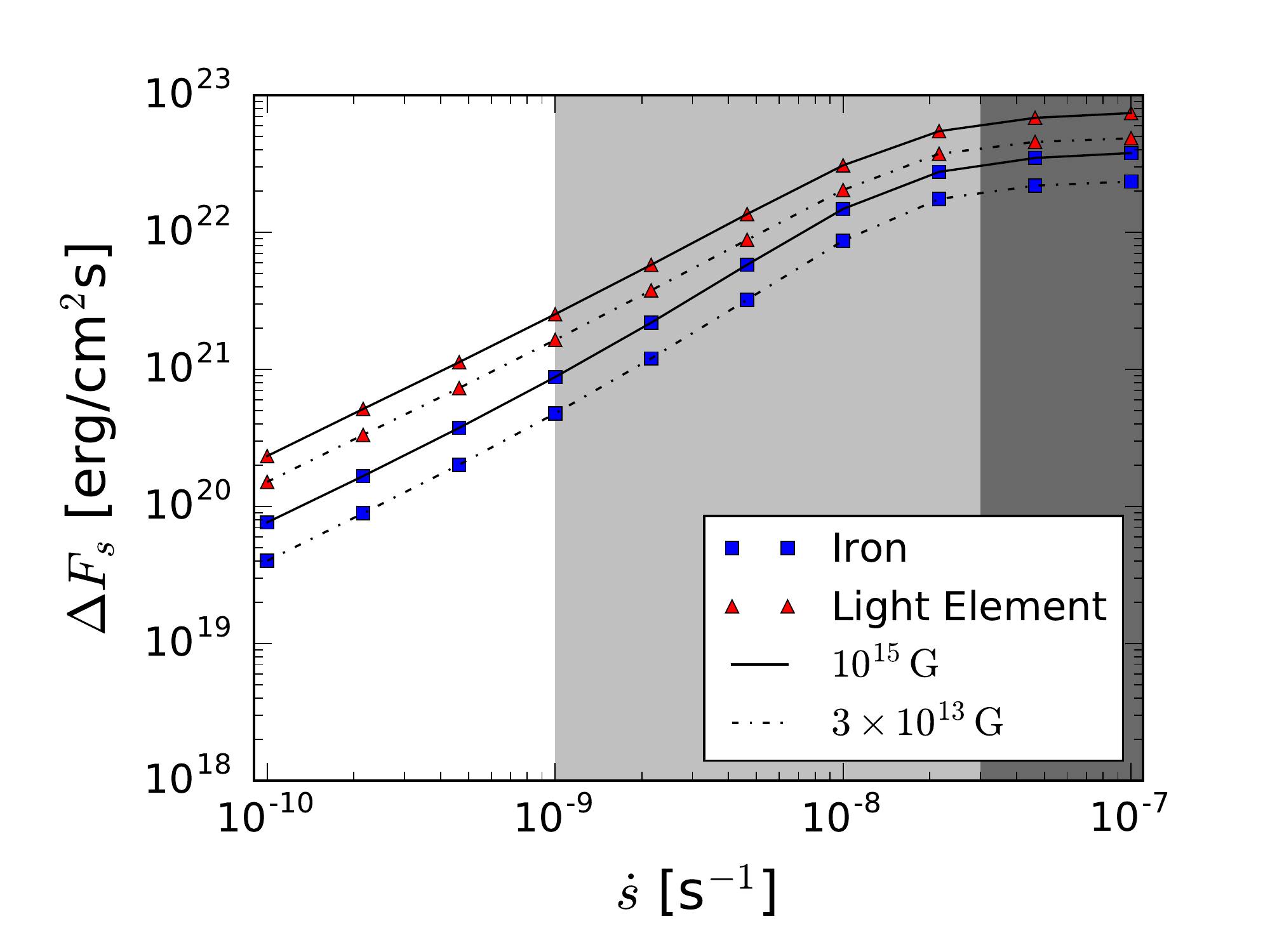} 
\end{tabular}
\caption{Surface radiation flux generated by the maximum possible mechanical 
dissipation $\dqh^{\max}=\sigma_{\max}|\dspl|$ (\Eq~\ref{eq:dqmax}).
Upper panel: the maximum dissipation occurs everywhere in the upper crust 
$\rho<10^{12}$~g~cm$^{-3}$. Lower panel: the maximum dissipation occurs 
in the entire crust $\rho<10^{14}$~g~cm$^{-3}$.
The magnetic field is assumed to be radial, which is the best possible configuration 
for maximizing the surface flux. 
The shaded region corresponds to shear rates exceeding $s_{\max}/t$ 
estimated in \Eq~(\ref{eq:smax}).
Shear rates in the darker region $\dspl>1$~yr$^{-1}$ would be able 
to sustain external magnetic twists against their resistive dissipation in the 
magnetosphere \citep{2009ApJ...703.1044B}.
 }
 \label{fig:mech}
 \end{figure}

This conclusion is illustrated by the numerical model assuming the maximum possible 
mechanical heating (Figure~\ref{fig:mech}). The model makes the most optimistic (and 
unrealistic) assumption that the crust flows everywhere with $\sigma=\sigma_{\max}$.
Even in this case, $\Fs$ can barely approach $10^{22}$~erg~s$^{-1}$~cm$^{-2}$, 
as long as $|\dspl|\ll 0.1$~yr$^{-1}$.

\subsection{Pumping crustal strain by Hall drift}\label{5.2}

Feeding the surface radiation flux by mechanical dissipation would become possible if 
the crust experiences an {\it oscillating} plastic flow with an effective $\Delta s\gg 100$. 
Large-amplitude oscillating shear could, in principle, be fed by the internal toroidal field 
energy without requiring a quick reduction of the internal $B_\phi$.
In particular, Hall waves in the crust is a possible driver of the oscillations \citep{2016arXiv160604895L}. 

The magnetic field evolves according to the equation
$\partial\bB/\partial t=-c\nabla\times\bE$. The electric field in the crust satisfies the 
generalized Ohm's law which expresses the balance of forces applied to the electron fluid,
\beq
\label{eq:el_bal1}
  \bE+\frac{\bv_e\times\bB}{c}+\frac{\nabla P_e}{e\,n_e}-\frac{m_e^\star{\mathbf g}}{e}
  =\frac{\bj}{\cond}, 
\eeq
where $\bj=(c/4\pi)\nabla\times\bB$ and $\cond$ is the electric conductivity.
In contrast to \Eq~(\ref{eq:el_bal}), here we included the resistive term $\bj/\cond$. 
This gives
\beq
\label{eq:el}
   \frac{\partial\bB}{\partial t}=\nabla\times \left[(\bv+{\mathbf v}_H)\times\bB
   -\frac{c}{\cond}\,\bj\right], \quad
    {\mathbf v}_{\rm H}=\frac{\mathbf j}{en_e}, 
\eeq 
where $\bv$ is the velocity of the ion lattice/liquid, and $\bvH=\bv_e-\bv$ is the velocity 
of the electrons relative to the ions (the Hall drift). As long as the ohmic term $c\bj/\cond$ 
is negligible, the magnetic field is frozen in the electron fluid moving with 
$\bv_e=\bv+{\mathbf v}_{\rm H}$. 
The ion motion $\bv\neq 0$ occurs in response to magnetic forces, not only 
in the liquid ocean but also in the solid crust, as a result of elastic or plastic deformations.
This motion can offset Hall drift. Previous numerical simulations of Hall drift 
in the crust used \Eq~(\ref{eq:el}) with $\bv$ set to zero, neglecting ion motion 
(e.g. \citealp{2009A&A...496..207P,2013MNRAS.434..123V,2016PNAS..113.3944G}).

For our purposes it is instructive to look at the force balance for ions,
\beq
\label{eq:ions}
  \bE+\frac{\bv\times \bB}{c}+\frac{Am_p{\mathbf g}}{Ze}
  -\frac{\nabla\cdot {\sigma}}{en_e}=0,
\eeq
where $Z$ and $A$ are the ion charge and mass numbers, and
$\sigma$ stands for ${\sigma}_{ik}$ --- the ion stress tensor. Using the expression 
for $\bE$ provided by \Eq~(\ref{eq:ions}) and taking into account that
$\nabla (A/Z)\parallel \nabla \Phi_g=-{\mathbf g}$, one finds 
\beq
    \frac{\partial\bB}{\partial t}=\nabla\times \left(\bv\times\bB
    -\frac{c\,\nabla\cdot\sigma}{en_e}\right).
\eeq
The second term in parenthesis determines the drift of the magnetic field relative to the ions.
A conservative upper limit on the ion stress tensor components is given by the 
Coulomb energy density, and is also comparable to the shear modulus of the crust $\mu$. 
Therefore, one can roughly estimate 
\beq
   \left|\frac{\partial\bB}{\partial t}\right|_{\rm H}^{\max}\sim \frac{c \mu}{en_e \ell^2}
   \sim \frac{c\,m_p v_{\rm sh}^2}{e Y_e \ell^2},
\eeq 
where $\ell$ is a characteristic scale of stress variations and 
$v_{\rm sh}=(\mu/\rho)^{1/2} \sim 10^8$~cm~s$^{-1}$ is the speed 
of shear waves sustained by the ion lattice; this speed
is approximately uniform throughout the solid crust (e.g. \citealp{1991ApJ...375..679S}). 
This gives an estimate for the maximum strain rate pumped by the Hall drift,
\beq
\label{eq:sH}
   \dot{s}_{\rm H}\simlt \frac{1}{B}\left|\frac{\partial\bB}{\partial t}\right|_{\rm H}^{\max}
     \sim \frac{10^{-3}{\rm ~yr}^{-1}}{Y_e\, B_{15}\, \ell_4^{2}}.
\eeq
The highest rate can be reached in the deep crust where $Y_e$ decreases to $\sim 0.1$. 
The rate $\dot{s}_{\rm H}$ can cause plastic flow with a comparable time-averaged strain 
rate $\dot{s}\sim\dot{s}_{\rm H}$. It is lower than needed for 
mechanical dissipation to keep the magnetar surface at $T_s\approx 4\times 10^6$~K.

Note also that the tension 
of magnetic field lines $\mu_B=B^2/8\pi$ exceeds the shear modulus of the upper crust 
$\mu\sim 10^{27}\rho_{11}$~erg~cm$^{-3}$, and $\mu$ practically vanishes in the ocean. 
This fact alone suggests that Hall drift cannot cause interesting deformations of the 
magnetic field in the upper layers. The presence of significant $\dot{s}_{\rm H}$ 
by itself does not imply significant field deformations, because it can be offset by the 
ion motion that limits the growth of shear stress.


\subsection{Intermittent mechanical dissipation}

The main conclusion of \Sects~\ref{5.1} and \ref{5.2} is that  
mechanical dissipation driven by internal evolution of the magnetic field in the 
star is too weak to sustain the observed persistent surface luminosity of magnetars.
Strong mechanical heating is only possible in an intermittent regime, where 
part of magnetic energy is suddenly dissipated due to an instability. The
instability can happen inside the crust (a thermoplastic wave or an avalanche
of failures driven by short Hall waves) or outside the star (a magnetospheric flare). 

In general, the efficiency of surface heating by mechanical dissipation is maximized 
when the dissipation takes place at a minimum depth, just below the liquid ocean. 
This naturally occurs when a strong high-frequency shear wave is launched from
the magnetosphere toward the crust, as expected in a powerful magnetospheric flare.
Therefore, we now focus on this more promising mechanism.

The magnetospheric wave damping somewhat increases the depth of the ocean 
by melting the crust, so that the heat deposition has to peak at the transition to the 
solid phase \citep{2015ApJ...815...25L}.
This heating occurs very quickly, on a timescale $\sim 10$~ms.
The  Alfv\'en waves excited by the flare create a 
train of $\sim 10$ strong oscillations of the crust, with a compressed and amplified strain, 
and produce a net plastic strain flow $\Delta s$ that can exceed $10$.
Most of the plastic dissipation occurs in a layer of thickness 
$\Delta z\sim 100$~m at a depth of a few hundred meters. 
This depth is found by balancing the wave energy deposited per unit area 
of the crust, $Q$, with the energy it takes to melt the layer,
\beq
\label{eq:melt}
   Q\sim \Delta z \int_0^{T_m} C_V dT,
\eeq
where $\Tm\approx 10^{9}\rho_{11}^{1/3}$~K is the melting temperature, and $C_V$ 
is the heat capacity; for instance,
$C_V\sim 4\times 10^{17}$~erg~cm$^{-3}$~K$^{-1}$ at $\rho=10^{11}$~g~cm$^{-3}$
and $T\approx \Tm$ (e.g. \citealp{2001MNRAS.324..725G,2015SSRv..191..239P}).
\Eq~(\ref{eq:melt}) determines the characteristic density at which the wave is damped;
it is comparable to $10^{11}$~g~cm$^{-3}$ for $Q\sim 10^{30}-10^{31}$~erg~cm$^{-2}$
and grows with $Q$. 

\begin{figure}[t]
\begin{tabular}{c}
\hspace*{-6mm}
\includegraphics[width=0.51\textwidth]{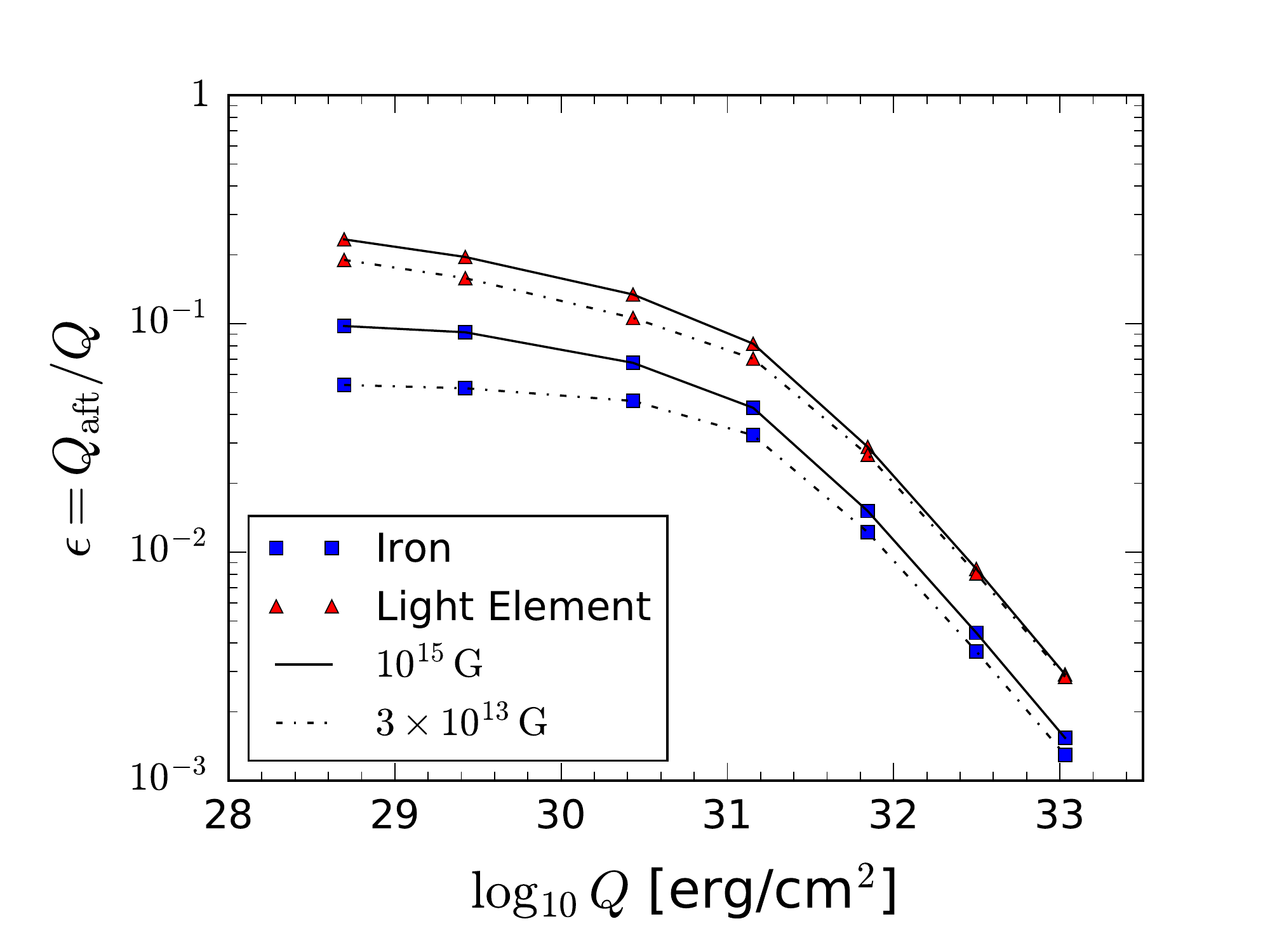} 
\end{tabular}
\caption{Efficiency $\eff=Q_{\rm aft}/Q$ of surface heating by plastic damping of 
Alfv\'en waves from a magnetospheric flare.
The efficiency is defined  as the fraction of the deposited energy that is 
radiated from the surface (rather than conducted to the core and lost to neutrino 
emission). It is shown as a function of the 
deposited energy per unit area of the crust $Q$. A radial magnetic field was assumed
in the calculations, which gives the maximum $\eff$.
 }
 \label{fig:trans}
 \end{figure}
\begin{figure}[t]
\begin{tabular}{c}
\hspace*{-6mm}
\includegraphics[width=0.51\textwidth]{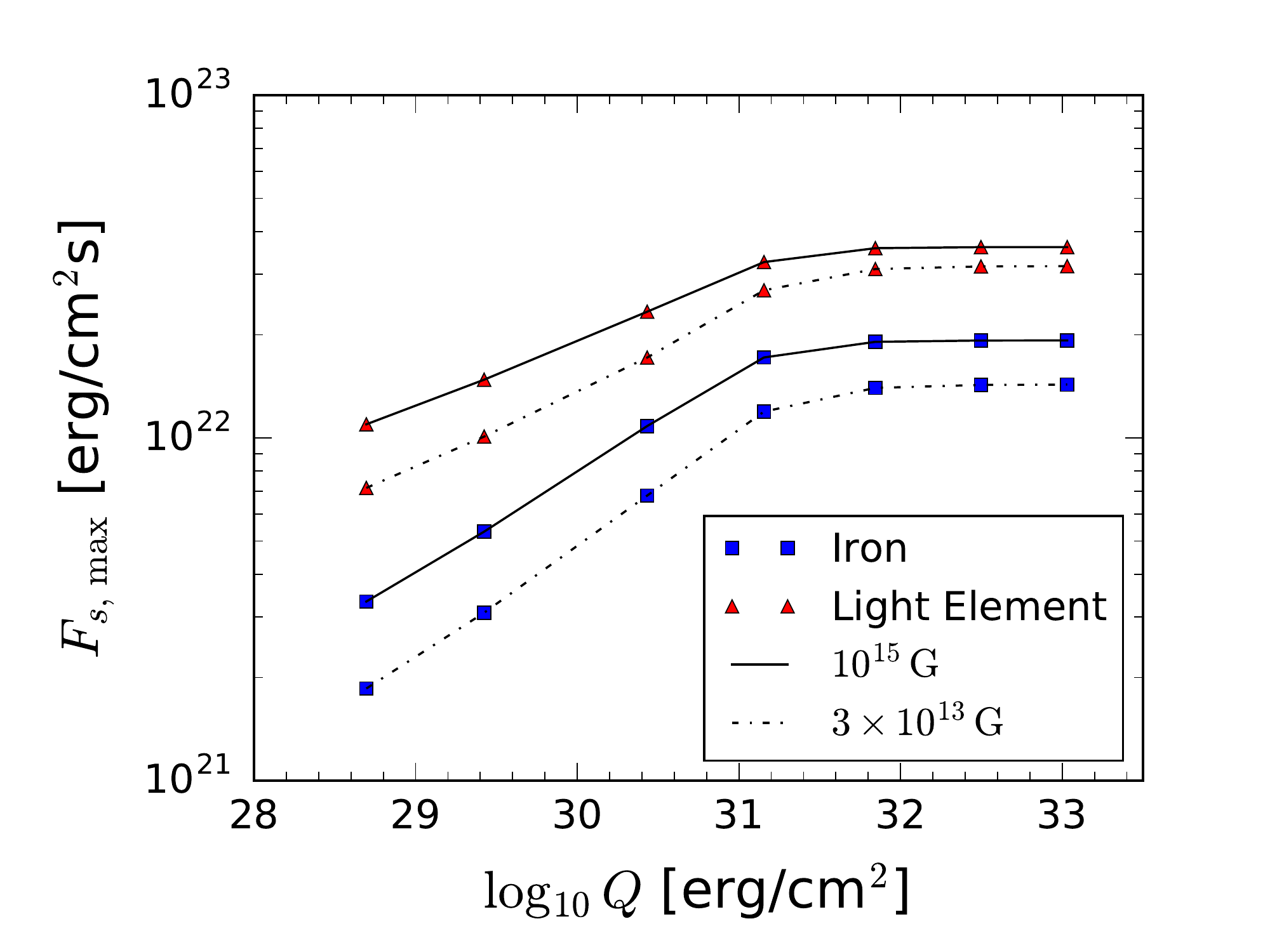} 
\end{tabular}
\caption{Peak flux of the surface radiation following the plastic damping of 
Alfv\'en waves. The peak lasts the cooling time (comparable to one year).
 }
 \label{fig:peak}
 \end{figure}

A fraction $\eff$ of the deposited heat $Q$ is gradually conducted from the deep melted 
ocean (where $T\approx \Tm$ immediately after the heating event) to the surface. 
This fraction is maximum when the magnetic field is approximately 
radial (vertical) in the ocean. We have calculated $\eff$ for this case using 
detailed time-dependent simulations of heat conduction and neutrino cooling.
The method of our calculations is similar to previous simulations of time-dependent
heat transfer in a neutron star crust (e.g. \citealp{2009ApJ...698.1020B,2009A&A...496..207P,2014MNRAS.442.3484K}) and described in \citet{2015ApJ...815...25L}.  

Figure~\ref{fig:trans} shows the result. When $B\sim 10^{15}$~G the afterglow efficiency 
$\eff=Q_{\rm aft}/Q$ can be approximated by the formula, 
\beq
   \eff\approx  \eff_0\,(1+2 Q_{31})^{-3/4},
\eeq
where $\eff_0\approx 0.1$ and 0.2 for iron and light element envelopes,
respectively.  A strong wave delivering energy $Q\gg 10^{30}$~erg~cm$^{-2}$ 
results in deep melting of the crust and deposits energy at large depths, which 
reduces $\eff$. Therefore, the afterglow energy radiated per unit area of the 
crust $Q_{\rm aft} = \eff Q$ saturates near  a few times $10^{30} {\rm ~erg~cm}^{-2}$,
slowly changing with $Q>10^{31}{\rm ~erg~cm}^{-2}$.

The peak flux of the surface afterglow is shown in Figure~\ref{fig:peak}. It is reached on 
the heat conduction timescale of the ocean, $t_c\sim 10^7$~s, and then gradually 
decays as the crust cools. The characteristic afterglow flux from the surface is 
$F_s\sim \eff Q/\tc$. Our calculations assumed a single flare, however, a similar result 
would be obtained if $N$ flares occur during time interval $t<\tc$, as long as $Q$ 
represents their cumulative energy deposition over the time $t_c$. The frequent flares
may have a 
slightly higher efficiency of surface heating, because of lower neutrino cooling, as each 
individual heating event $Q/N$ is weaker at large $N$ and has a lower peak temperature.
At $N\gg 1$, the heating approaches the quasi-steady regime with the self-consistent 
$\zm$ that was considered in Section~\ref{5.1}.


\section{Ohmic dissipation in the crust}\label{ohm}

Magnetars may have strong non-potential magnetic fields stored in the crust and 
sustained by electric currents, which satisfy the relation $(4\pi/c)\bj=\nabla\times\bB$. 
Ohmic dissipation tends to convert the stored energy of non-potential field to heat.
The rate of this process is controlled by the electric conductivity.

\subsection{Electric conductivity}

The electric conductivity of the crustal material is related to its thermal conductivity, 
as both charge and heat are transported by the electrons. The conductivities are controlled 
by the electron interaction with atomic nuclei (which form the lattice in the solid phase or 
the strongly coupled Coulomb liquid in the ocean) and by the magnetic field.
The conductivity tensor $\cond_{ik}$ in the magnetized crust is described by three 
components: $\cond_\parallel$ (conductivity parallel to the magnetic field), $\cond_\perp$ 
(perpendicular to the field), and the Hall component $\cond_{\rm H}$ (the antisymmetric 
off-diagonal component of the tensor $\cond_{ik}$, see e.g. \citealp{1960ecm..book.....L}). 
Detailed calculations of $\cond_{ik}$ for densities, temperatures, 
and magnetic fields relevant to neutron starts are found in \citet{1999A&A...351..787P}.

For a given electric current density $\bj$, the electric field $\bE$ can be found by 
inverting the relation $j_i=\cond_{ik} E^k$. It is useful to express the electric current 
as the sum of components parallel and perpendicular to $\bB$, $\bj=\bj_\parallel+\bj_\perp$. 
Then the rate of ohmic heating is given by
\beq
  \dqohm=\bE\cdot\bj=\frac{j_\parallel^2}{\cond_\parallel}
            + \frac{j_\perp^2}{\cond_\perp^{\rm eff}},
\eeq
where 
\beq
   \cond_\perp^{\rm eff}=\cond_\perp+\frac{\cond_{\rm H}^2}{\cond_\perp}
   \approx \frac{\cond_{\rm H}^2}{\cond_\perp}
\eeq
is the effective conductivity perpendicular to $\bB$. Electron collisions resist 
$\bj_\parallel$ and help conduct $\bj_\perp$ with a non-zero component along $\bE$. 
Without collisions, $\bj_\perp$ would be the pure drift current proportional to 
$\bE\times\bB$, which does not contribute to ohmic dissipation $\bE\cdot\bj$.

\begin{figure}[t]
\begin{tabular}{c}
\hspace*{-6mm}
\includegraphics[width=0.52\textwidth]{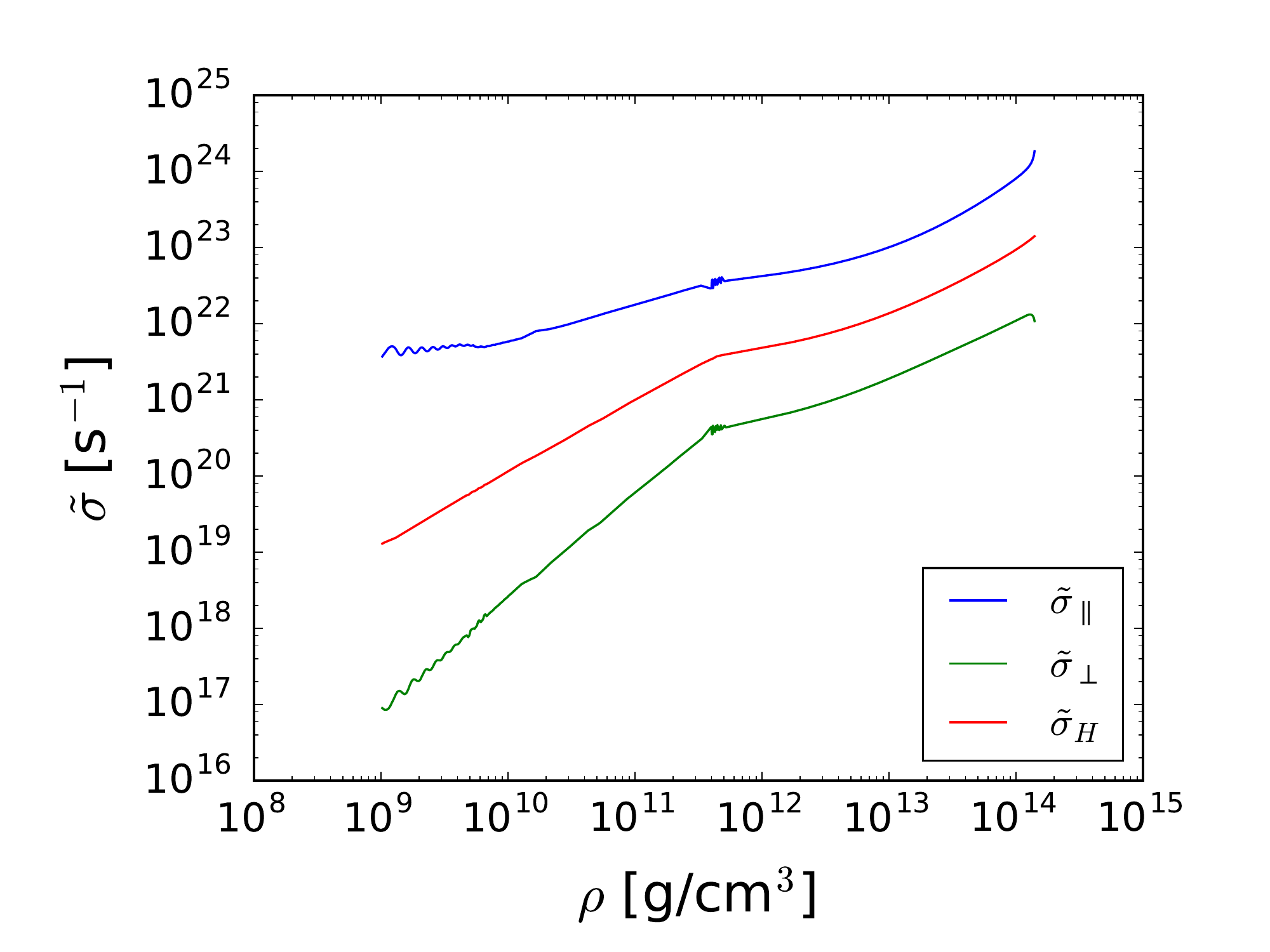} 
\end{tabular}
\caption{Components of the conductivity tensor in the crust with a 
steady temperature profile that sustains $T_s=4\times 10^6$~K. 
The temperature profile was calculated assuming a radial magnetic field 
$B=3\times 10^{14}$~G and an iron envelope. 
Temperature $T\approx 8.7\times 10^{8}$~K is approximately uniform 
in the region of $\rho>10^{10}$~g~cm$^{-3}$ ($T$ steeply decreases toward 
the surface in the blanketing envelope). In the presence of a heat source
in the crust at depth $z_h$, the curves can only be used at $z<z_h$.
}
 \label{cond}
 \end{figure}

The components of the conductivity tensor obey the following relations 
(e.g. \citealp{1990A&A...229..133H}),
\beq
   \cond_\perp=\frac{\cond_\parallel}{1+a^2}, \qquad \cond_{\rm H}=a\cond_\perp,
\eeq
where $\cond_\parallel=(e^2n_e/m_e^\star)\tau_0$ is related to the electron collision 
time $\tau_0$, $m_e^\star$ is the electron inertial mass, and $a=\tau_0\, eB/m_e^\star c$ 
is the magnetization parameter. For magnetar fields $a\gg 1$, and then 
$\cond_\perp^{\rm eff}\approx\cond_\parallel$. Therefore, one can use the simple 
equation,
\beq
    \dqohm=\frac{j^2}{\cond}, \qquad \cond\approx\cond_{\parallel}\approx\cond_\perp^{\rm eff}.
\eeq
Figure~\ref{cond} shows $\cond_\parallel$, $\cond_\perp$, $\cond_{\rm H}$
for a steady temperature profile with $T_s=4\times 10^6$~K and an iron envelope.
In the main region of interest, where $\rho=10^9-10^{11}$~g~cm$^{-3}$, 
$\cond\sim 10^{22}$~s$^{-1}$. Note also that in the region where heating occurs the 
conductivity will be reduced, because of the local increase in temperature.

\subsection{Dissipation of electric currents in the crust}

The timescale for dissipating electric currents that sustain variations $\delta B$
on a scale $\ell$ is 
\beq
\label{eq:tohm}
   \tohm= \frac{4\pi\tilde{\sigma} \ell^2}{c^2}
         \approx 4\times 10^4\, \tilde{\sigma}_{22}\,\ell_{\rm km}^2  {\rm~yr}.
\eeq
This timescale in the upper crust may be comparable to the magnetar age of 1-10~kyr
if the field varies on a scale $\ell\sim 0.3$~km. The corresponding electric current,
\beq
    j\sim \frac{c}{4\pi}\,\frac{\delta B}{\ell},
\eeq
produces the heating rate
\beq
   \dqohm
   \sim \frac{(\delta B)^2}{4\pi \tohm}
   \sim 6\times 10^{18}\,(\delta B_{16})^2\,\ell_{\rm km}^{-2}\,\cond_{22}^{-1}
   {\rm ~erg~s}^{-1}{\rm cm}^{-3}.
\eeq
A minimum heating rate $\sim3\times 10^{19}$~erg~s$^{-1}$~cm$^{-3}$ capable  
of sustaining $T_s\sim 4\times 10^6$~K \citep{2014MNRAS.442.3484K}, can be 
achieved if the field varies on a small scale $\ell\sim 0.3$~km and these variations 
are large, $\delta B\sim 10^{16}$~G, which requires an ultrastrong field, $B>10^{16}$~G. 
Such crustal fields were invoked by \citet{2011ApJ...741..123P} to explain the surface 
luminosity of magnetars. Their model of AXP~1E~2259+586 assumes 
a toroidal magnetic field $B=2.5\times 10^{16}$~G hidden in the middle of the crust,
which drops toward the core and toward the stellar surface on a scale comparable 
to 0.3~km. Similar configurations with weaker fields evolving due to the combined effects 
of Hall drift and ohmic dissipation were simulated by \citet{2009A&A...496..207P} 
and \citet{2013MNRAS.434..123V}.
They argued that the magneto-thermal evolution of crustal fields can explain 
the observed properties of a broader class of neutron stars, not only magnetars.

The requirements to the ohmic heating model are illustrated in more detail 
by the following calculation.
Let $z_h$ be the characteristic depth where the ohmic heating occurs. 
The corresponding heated volume is $V=\Delta z\,A$, where $\Delta z$
is the thickness of the heated layer and $A\simlt 10^{13}$~cm$^2$ is its area.
Suppose this heating sustains the observed surface temperature 
$T_s\approx 4\times 10^6$~K. The heat transfer equation determines 
the required heating rate $F_h=\Delta z\,\dqohm$ and $T(z_h)$.
The calculation is simplified if we use the approximation of a thin 
heated layer $\Delta z\ll z_h$ (Section~\ref{balance}). Then the required $F_h$ 
is independent of $\Delta z$, and a realistic $\Delta z\simlt z_h$ only enters 
at the final step when evaluating the required $\dqohm=F_h/\Delta z$.
The obtained temperature $T(z_h)$ determines the conductivity $\cond(z_h)$, 
and one can find $|\nabla\times\bB|=(4\pi/c)(\cond\dqohm)^{1/2}$ 
that is required in the heated region.

\begin{figure}[t]
\begin{tabular}{c}
\hspace*{-6mm}
\includegraphics[width=0.52\textwidth]{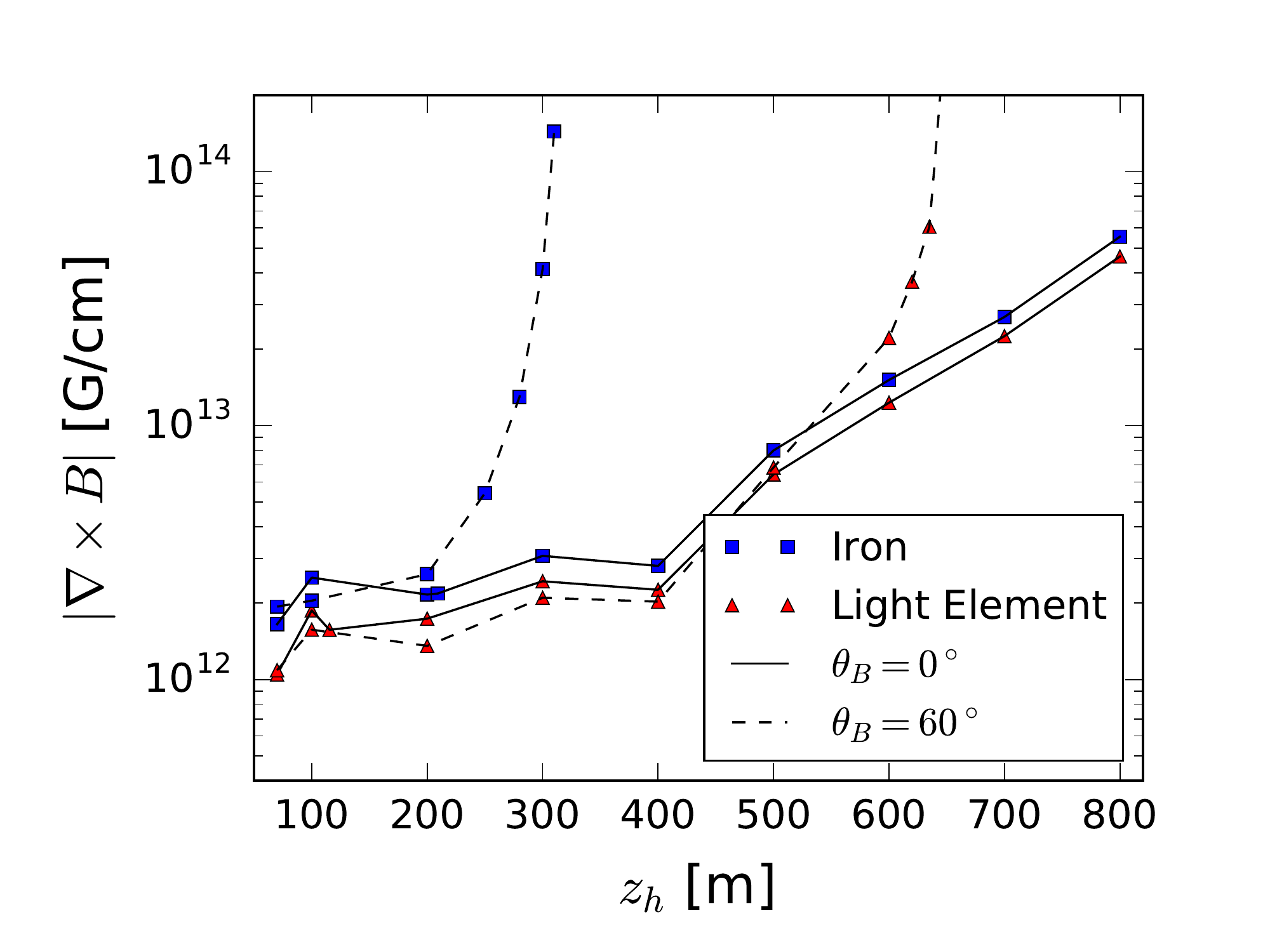} 
\end{tabular}
\caption{The required $|\nabla\times\bB|=(4\pi/c)j$ in the ohmically heated layer
if the heating is to sustain $T_s=4\times 10^6$~K. The required $|\nabla\times\bB|$
depends on the position of the ohmic heater $z_h$. The thickness of the heated 
layer was estimated as $\Delta z=z_h/2$. Magnetic field $B=10^{15}$~G is 
assumed and two cases are shown: $\thB=0$ (radial field) and $\thB=60^{\rm o}$.
}
\label{fig:ohm}
\end{figure}

The result of this calculation is shown in Figure~\ref{fig:ohm} as a function of $z_h$,
assuming $\Delta z=z_h/2$. One can see that $|\nabla\times\bB|>10^{12}$~G~cm$^{-1}$ 
is required by a successful ohmic heating model, which corresponds to 
field variations $\delta B\simgt 10^{16}$~G on a 0.1-km scale.
If the field is dominated by a non-radial component, heat conduction across the envelope 
is reduced; then for a heater located deep below the surface it becomes impossible to 
sustain $T_s=4\times 10^6$~K regardless of the ohmic power $\dqohm$.
The required temperature at $z_h$ becomes so high that neutrino losses prevent 
from reaching it, leading to the runaway of the required $F_h$ and $|\nabla\times\bB|$.

The ultrastrong crustal fields invoked by the ohmic heating model imply the following 
special feature.
Magnetic energy density $B^2/8\pi\approx 4\times 10^{30}B_{16}^2$~erg~cm$^{-3}$ 
exceeds the crustal shear modulus $\mu\sim 10^{28}\rho_{12}$~erg~cm$^{-3}$, 
and hence the maximum elastic stress $\sigma_{\max}\sim 0.1\mu$ 
is far below the magnetic stress. In this situation, the
crust should be viewed as an incompressible stratified liquid, with practically zero
tolerance to unbalanced shear stresses. In particular,
in an axisymmetric configuration, the toroidal component 
of the Lorentz force cannot develop, ${\mathbf e}_\phi\cdot(\bj\times\bB)/c\approx 0$. 
This condition implies that the poloidal current $\bj_p$ 
is nearly parallel to the poloidal magnetic field $\bB_p$,
\beq
\label{eq:jp}
   \bj_p\times\bB_p\approx 0.
\eeq
As long as the strong currents are confined to the crust,  
\Eq~(\ref{eq:jp}) requires that the current-carrying field lines are 
also closed below the stellar surface. 

Another special feature of this configuration is that the effect of Hall drift is limited 
(cf. the end of Section~5.2).
Like the magnetized liquid in the ocean, the magnetically dominated solid crust 
should follow the field in its relaxation to the lowest MHD equilibrium accessible 
through horizontal plastic shear motions
(vertical motions are constrained by the stable stratification of the crust). 
The class of such constrained MHD equilibria is rather broad 
\citep{2013MNRAS.433.2445A}.

\subsection{Ohmic dissipation in current sheets}

Currents sheets with thickness $\ell\ll 0.1$~km would produce a high local dissipation 
rate $\dqohm=j^2/\cond$. The immediate result is the growth of thickness $\ell(t)$ on 
the timescale $\tohm$ given by \Eq~(\ref{eq:tohm}). This limits the energy dissipated 
at given $\ell$ before the current sheet doubles its thickness.
The magnetic energy that is released by a current sheet of area $A$ and thickness $\ell$
sustaining a field jump $\delta B$ is 
\beq
E_{\rm diss} \sim A\,\ell\,\frac{(\delta B)^2}{8\pi}
\sim 4\times 10^{45}\,A_{12}\,\ell_{\rm km}\,(\delta B_{15})^2.
\eeq
Feeding the magnetar surface luminosity during its lifetime $t\sim 10^{11}$~s
requires large heat $E_h\sim \eff^{-1} t \Lum_s\sim 10^{46}\eff^{-1}$~erg, which
implies tapping into magnetic energy in a large fraction of the crust volume. 
Therefore, formation of thin current sheets by itself is insufficient to explain the 
surface luminosities of magnetars. The high rate of ohmic dissipation could only
be sustained if some process prevents the current sheet from thickening and also 
advects magnetic energy into it, feeding its dissipation power. 

Hall drift is a process that could in principle do this. 
In particular, consider a horizontal field $B_y$ which varies in the orthogonal horizontal 
direction $x$;\footnote{In the presence of other components of the magnetic field, 
     the current sheet formation is less efficient \citep{2004MNRAS.347.1273H}; 
     therefore we focus here on the simple and most optimistic configuration $B_y(x)$.}
the vertical $z$-axis is chosen along the electron density gradient $\nabla n_e$. 
As long as resistivity is neglected, the Hall drift of $B_y$ is described by 
\beq
   \frac{4\pi e}{c}\frac{\partial B_y}{\partial t}=\frac{d}{d z}\left(\frac{1}{n_e}\right) B_y\,\frac{\partial B_y}{\partial x}.
\eeq
Its behaviour is similar to a non-linear wave described by Burger's equation, as 
discussed by \citet{2000PhRvE..61.4422V}. The profile of $B_y(x)$ can continue to 
steepen until resistivity becomes important and the magnetic diffusivity offsets the 
steepening. Then a current sheet of a small thickness $\ell$ will be sustained.

The resulting energy dissipation rate is controlled by the speed of Hall drift 
that advects magnetic energy toward the current sheet. This rate is formed outside 
the current sheet and independent of its thickness $\ell$. 
Thus, tapping into magnetic energy stored in a large volume anyway relies 
on electric currents far from the current sheet.
The large-scale Hall drift transports energy slowly, in particular in the deep 
dense crust that takes most of the volume and stores most of the magnetic energy. 

The fastest energy transport due to Hall drift would occur in small-scale Hall
waves propagating along the magnetic field lines with the group speed
$v_{\rm gr}=cBk/2\pi en_e$ where $k$ is the wavenumber \citep{1992ApJ...395..250G}. 
However, very short waves are ohmically damped. The shortest waves that can 
propagate an interesting distance $H\simgt 10^4$~cm have 
\beq
   k_{\max}\sim \frac{\cond B}{en_eH},
\eeq
and their energy transport time is 
\beq
  t_{\min}\sim \frac{H}{v_{\rm gr}^{\max}}
\sim 10\, H_4^2\,n_{e,36}^2 B_{15}^{-2}\cond_{24}^{-1} {\rm~yr}.
\eeq
A mechanism generating short Hall waves could lead to fast energy transport across
the crust and assist ohmic or mechanical dissipation; this scenario is investigated
in \citet{2016arXiv160604895L} and also found incapable of sustaining the surface 
luminosity of persistent bright magnetars.

Another possibility for creating current sheets was considered by 
\citet{2001ApJ...561..980T}. In their scenario, magnetar starquakes produced 
crustal fractures with localized shear. Shear localization along a fault surface would 
create a jump of the (tangential) magnetic field --- a current sheet. 
This could occur if the crust breaks and slides along a magnetic flux surface ---
otherwise the transverse field suppresses such sliding \citep{2012MNRAS.427.1574L}.
It was proposed that the current sheets induced by crustal fractures could quickly 
dissipate a large magnetic energy through reconnection \citep{2001ApJ...561..980T,2002ApJ...580L..69L}.

This scenario is however problematic. Strong magnetic fields may exist when they are 
rooted in the deep crust, which keeps the field in place.
The current sheet created by localized shear is immersed 
in a guide field that is frozen in the lower crust and therefore cannot be moved 
out of the sheet, inhibiting reconnection.\footnote{For a similar reason the current 
       sheet hugging the closed magnetosphere of a rotation-powered pulsar is stable. 
       Direct plasma simulations of pulsar magnetospheres show fast reconnection 
       only in the equatorial part of the current sheet outside the light cylinder, where 
       a guide field is absent \citep{2014ApJ...795L..22C,2015ApJ...801L..19P,2015MNRAS.448..606C}.}
The current sheet will simply thicken 
with time due to resistive magnetic diffusion, and ohmic dissipation will 
become slow before tapping into the larger reservoir of magnetic energy. 

A network of $N\gg 1$ fractures occupying a large region of scale $L$ 
would reduce the distance between the multiple current sheets to $L_0=L/N$. 
However, it would also reduce the field jump $\delta B\sim B/N$ 
in each sheet, resembling a staircase with many small stairs. 
Only a small magnetic energy converts to heat before ohmic 
dissipation washes out the ``stairs'' and makes the field profile smooth.
This energy may be estimated as 
\beq 
   E_h\sim V\,\frac{(\delta B)^2}{8\pi}\sim N^{-2}\,V\,\frac{B^2}{8\pi},
\eeq
where volume $V\simlt 10^{18}$~cm$^3$ does not exceed the volume of the crust.
The dissipation timescale for this small energy is short, 
$\tohm=4\pi\cond L_0^2/c^2\sim 4\pi\cond L^2/N^2c^2$.
However, dissipation of the main magnetic energy can only occur on a long ohmic 
timescale that corresponds to scale $L$ comparable to the size of the magnetic 
energy reservoir. In summary, we do not find any scenario for
efficient crustal heating by current sheets.


\section{External heating}

Magnetar surface can be heated by relativistic magnetospheric particles. 
Clear evidence for magnetospheric activity is provided by hard X-ray observations: 
persistent magnetars show a strong nonthermal component in their spectra,
peaking at photon energies $E>100$~keV 
\citep{2008arXiv0810.4801K,2010PASJ...62..475E}. The power released in the 
magnetosphere exceeds the surface luminosity $\Lum_s$, 
and partial reprocessing of this power may be sufficient to feed $\Lum_s$. 

The source of hard X-rays was identified as a decelerating outflow of copious 
$e^\pm$ pairs in the closed magnetosphere
\citep{2013ApJ...777..114B,2013ApJ...762...13B,2014ApJ...786L...1H,2015ApJ...807...93A}.
The $e^\pm$ fountain forms near the neutron star and radiates the observed hard 
X-rays at several stellar radii before reaching the top of the closed magnetic loop and 
annihilating there. The model successfully fitted the variation of the observed spectrum 
with rotational phase, and the fits determined the location of the $e^\pm$ fountain, 
in particular in 1RXS J1708-4009 and AXP~1E~1841-045. The fountain
typically operates on 1-10\% of magnetic field lines emerging from the star,
which form a twisted bundle carrying electric current $\bj=(c/4\pi)\nabla\times\bB$;
the observed activity is the result of electric discharge in this ``j-bundle.''

These results imply
that the hard X-ray emission is 
directed away from the star and cannot heat its surface. However, a significant fraction 
of the primary {\it particles} created by the discharge near the star are expected to 
flow toward the surface.\footnote{The energy flow from the discharge zone toward 
     the star is carried by relativistic particles rather than photons. 
     The main radiative process for the particles is resonant scattering of soft X-rays, 
     and in the ultrastrong field near the star it gives so energetic photons that they 
     immediately convert to $e^\pm$ pairs \citep{2007ApJ...657..967B}.
     In contrast, particles that flow away from the star and reach $B<10^{13}$~G   
     eventually radiate almost all their energy through resonant scattering.
     }
These particles must bombard the surface and heat it, forming a hot spot at the 
footprint of the j-bundle.

\begin{figure}[t]
\hspace*{-0.5cm}
\includegraphics[width=0.53\textwidth]{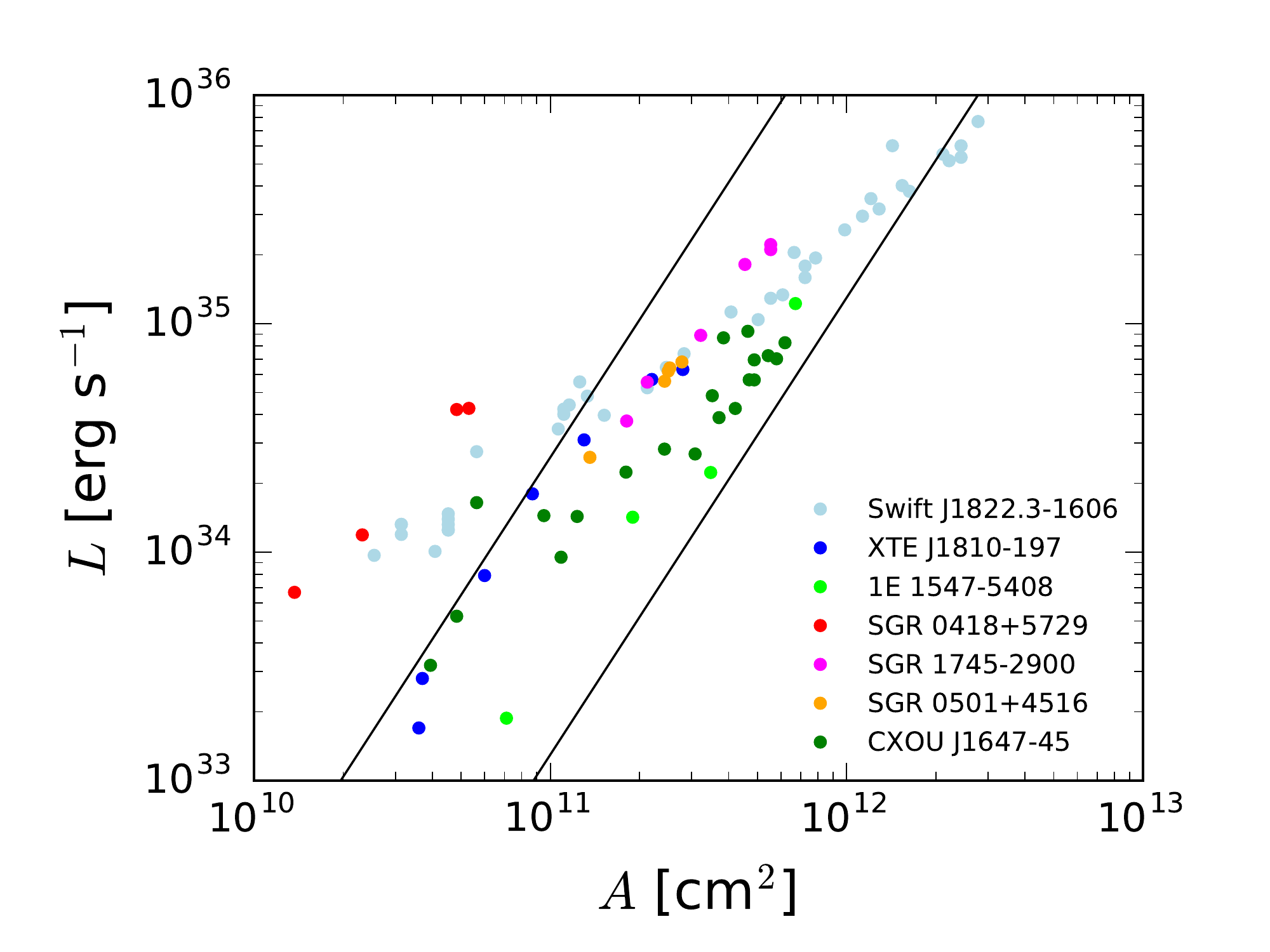} 
\caption{The evolution of hot spots observed on transient magnetars following 
their outbursts. The spot shrinks (its area $A$ decreases) and becomes dimmer 
(its luminosity $\Lum$ decreases) with time, forming tracks on the $A$-$\Lum$ plane.
The theoretical prediction is shown by the strip between the two lines, 
$\Lum\sim 1.3 \times 10^{33} K\,A_{11}^2$~erg~s$^{-1}$, where $K=B_{14} \Phi_9 \psi$
\citep{2009ApJ...703.1044B}.
The value of $K$ depends on the discharge voltage $\Phi\sim 10^9-10^{10}$~V, the 
twist angle $\psi\sim 1$, and the characteristic surface magnetic field $B$.
The strip shown in the figure corresponds to $1<K<20$,
however a broader range is possible, and $K$ may evolve during the outburst.
Data for SGR 1745-2900 are from \citet{2015MNRAS.449.2685C};
CXOU J1647-45 from \citet{2011ApJ...726...37W} and \citet{2013ApJ...763...82A};
Swift J1822.3-1606 from \citet{2012ApJ...754...27R};
SGR 0418+5729 from \citet{2010MNRAS.405.1787E};
SGR 0501+4516 from \citet{2009MNRAS.396.2419R};
XTE J1810-197 from \citet{2007Ap&SS.308...79G};
1E 1547-5408 from \citet{2008ApJ...676.1178H} and \citet{2010PASJ...62..475E}.
The distance to 1E~1547-5408 was changed to 4~kpc following  
\citet{2010ApJ...710..227T} and \citet{2007ApJ...667.1111G}.
 }
 \label{fig:spots}
 \end{figure}

Strong observational evidence for external heating exists for transient magnetars.
A canonical transient magnetar, e.g. \XTE, shows an outburst followed by a decay 
of emission on a timescale of months to years, returning to the quiescent state
\citep{2007Ap&SS.308...79G}. The outburst results from a shear motion of 
the magnetar surface twisting the external magnetosphere, which is followed by 
gradual untwisting on the resistive timescale. The timescale is regulated by the 
discharge voltage $\Phi\sim 10^{10}$~V that sustains the magnetospheric current $\bj$.
Electrodynamics of untwisting requires that the current becomes localized on a fraction 
of magnetic field lines, forming the j-bundle, and this fraction slowly shrinks with time 
\citep{2009ApJ...703.1044B}.  As the j-bundle shrinks so does its hot footprint.
Figure~\ref{fig:spots} summarizes observations of shrinking hot spots
in seven transient magnetars. The observed evolution of the spot area $A$ and 
luminosity $\Lum$ agrees with the special trend predicted by the untwisting 
magnetosphere model:
$A$ and $\Lum$ decrease with time. The slope of the $\Lum$-$A$ relation 
(controlled by the behavior of $\Phi$) varies between 1 and 2, in the theoretically 
expected region of the $\Lum$-$A$ plane.
The typical timescale of this evolution --- months to years ---
is also consistent with theoretical expectations, although there are outliers that 
require a more detailed modeling.

The predicted and observed localization of external heating 
in transient magnetars suggests that this mechanism does not dominate
$\Lum_s$ in persistent magnetars, as most of their surface emission apparently
comes from a large area comparable to $4\pi R^2$.


\section{Discussion}

The observed surface luminosity of persistent magnetars 
$\Lum_s\approx 10^{35}$~erg~s$^{-1}$ is a challenge to magnetar theory. 
Energy transport from the core heated by ambipolar diffusion
is an attractive scenario, which lead \citetalias{1996ApJ...473..322T} to propose
an explanation for $\Lum_s\approx 10^{35}$~erg~s$^{-1}$: it corresponds to
the highest core temperature that ambipolar heating could sustain against neutrino cooling. 
We find, however, that this scenario faces the following problem. Even in the best
case of a magnetar with a light-element envelope,
$\Tc\simgt 6\times 10^8$~K is required (Figure~\ref{fig_Tc}). Although ultrastrong 
magnetic fields can drive a fast ambipolar drift that generates a huge heating rate, we 
find that such hot cores have lifetimes shorter than the typical magnetar age 
(Figure~\ref{fig:Tc}), as long as the typical wavenumbers of the variation of $\bB$ in the 
core satisfy the plausible assumption $2\pi/\kk\simlt 20$~km.
The lifetime is short because the ambipolar drift is fast in the hot core. It is not slowed 
down by the induced pressure gradients in a compressive drift and is only 
limited by the p-n friction, which is modest at high temperatures. Assuming stronger 
magnetic fields helps increase the energy reservoir available for dissipation, however it also 
accelerates its dissipation, with enormous heat promptly released and lost to neutrino
emission. The hot stage $\Tc>6\times 10^8$~K becomes particularly short if the core becomes 
superfluid at this stage, as the transition to superfluidity both speeds up the ambipolar 
drift and enhances neutrino cooling.

The issue of short lifetime could be resolved if ambipolar drift is intermittent, 
which would allow the magnetar to enter ``ice ages'' between hot periods.
This would help explain the $1-10$~kyr ages of currently observed hot magnetars.
Objects classified as ``persistent'' after 4 decades of observations may not be truly 
persistent on longer timescales; their appearance may dramatically change over centuries. 
The surface luminosity would respond to changes in the core heating on the thermal 
conduction timescale, which is comparable to a few years.
Note that the reduced duty cycle of magnetar activity would imply a large number of 
undetected quiescent objects. Then the inferred magnetar population is increased from 
10-20\% to more than half of all neutron stars with age less than 10~kyr. 
Evidence for the dormant population is provided by the growing
number of transient magnetars.  They are discovered in their outbursts of activity, 
which are followed by the decay to the quiescent state.

It is unclear whether heating of the core can become intermittent due to
complicated dynamics of the magnetic field. 
The dynamics may be influenced by current sheets, which are naturally
created by ambipolar diffusion (\Sect~\ref{ambipolar} and Appendix~\ref{app}).
Three-dimensional global simulations of ambipolar drift may clarify the
possibilities and limitations  for variable core heating.

An alternative location for the internal heat source is the crust surrounding 
the liquid core. This possibility became popular in recent years 
(e.g. \citealp{2006MNRAS.371..477K,2011ApJ...727L..51P,2014ApJ...794L..24B}), 
and we have examined it here in some detail. 
Two mechanisms can heat the crust: mechanical dissipation and ohmic dissipation.
The dissipative shear deformations can be triggered by the slowly evolving magnetic 
field inside the star. However, we find that even with most optimistic assumptions, this 
scenario can hardly sustain the observed surface luminosity of persistent magnetars.
We have calculated upper limits on mechanical heating that result from two 
general constraints: (1) the mechanical heating must occur in the solid phase below 
the deep melted ocean, and (2) the heating rate is proportional to the shear stress, 
which cannot exceed $\sigma_{\max}\sim 0.1\mu$, where $\mu$ is the shear modulus
of the crustal material. Mechanical heating is also proportional to the crustal shear rate 
$\dot{s}$. The maximum average $\dot{s}$ over the magnetar lifetime fails to generate 
the observed surface luminosity $\Lum_s\approx 10^{35}$~erg~s$^{-1}$. Therefore,
we have also considered the possibility of {\it oscillating} plastic shear driven by 
crustal Hall waves and have shown that it also obeys an upper limit, which
cannot sustain the observed $\Lum_s$ over the magnetar lifetime (\Sect~\ref{5.2}). 
This mechanism can, however, explain the intermittent heating observed in transient 
magnetars (see \citet{2016arXiv160604895L}).

Ohmic heating approaches the needed rate only for extreme magnetic configurations 
with crustal fields $B>10^{16}$~G varying on a scale of 100~m (Figure~\ref{fig:ohm}). 
For instance, an ultrastrong toroidal field stored in the crust can be considered as an 
ohmic heater \citep{2011ApJ...741..123P}. However, it is unclear how so
energetic magnetic torus could form and 
remain confined to the crust of a nascent magnetar; such configurations were 
not seen among calculated stable MHD equilibria \citep{2009MNRAS.397..763B}.
We have further explored the possibility of crustal ohmic heating by localized current 
sheets envisioned by \citet{2001ApJ...561..980T} and \citet{2002ApJ...580L..69L}.
We found no way for the crustal current sheets to efficiently dissipate the magnetic 
energy that would explain the observed surface luminosities.

The difficulties with finding a compelling internal heating mechanism suggest the 
possibility that magnetars are heated as a result of their magnetospheric activity.
In particular, magnetospheric flares create strong intermittent dissipation in the crust. 
The flares launch powerful Alfv\'en waves \citep{2013ApJ...774...92P} 
which induce plastic flow in the crust and dissipate in $\sim 10$~ms
\citep{2015ApJ...815...25L}. This impulsive heating occurs immediately below the 
melted ocean, and heat conduction from this region sustains a high surface 
temperature for $\sim 1$~yr with a relatively high efficiency $\eff$ (Figure~\ref{fig:trans}). 
Repeated flares could keep the magnetar surface hot for a longer time.
In this picture, $\Lum_s\sim 10^{35}$~erg~s$^{-1}$ requires an average power 
released in the magnetospheric flares $\Lum_f\sim 10^{36}$~erg~s$^{-1}$.
Curiously, this $\Lum_f$ is comparable to the persistent nonthermal luminosity 
estimated from the hard X-ray observations of persistent magnetars.
 
A flare of total energy $E_f$ produces surface afterglow with energy $\Eaft=\eff\, \fw E_f$,
where $\fw$ is the energy fraction given to the Alfv\'en waves damped in the crust.
The fraction $1-\fw$ is promptly radiated away during the flare, and the ratio of the 
energies radiated in the prompt phase and its crustal afterglow is 
\beq
   \frac{\Eaft}{\Eprompt}=\frac{\eff\,\fw}{1-\fw}.
\eeq
If the magnetospheric flares occur much more frequently than once per year, the 
afterglow luminosity becomes quasi-steady. For instance, flares with 
$E_f\sim 10^{42}$~erg and a rate of 30~yr$^{-1}$ would sustain a surface 
luminosity $\Lum_s\sim 10^{35}~(\eff/0.1)\, \fw$~erg~s$^{-1}$. 
Each flare could involve a strong deformation of a ``flux rope'' carrying a 
fraction of the stellar magnetic flux. A large number of such localized
flares could occur in a complicated magnetic field, with many twisted flux ropes. 
A problem with this scenario is that the high flare rates are not observed with 
current instruments. Most of them would need to be hidden by assuming that their 
prompt phase is ``dark'': $1-\fw\ll 1$, so that most of the released magnetic energy 
goes to the excitation of Alfv\'en waves.

Finally, magnetars must be heated by relativistic magnetospheric particles bombarding
the stellar surface. This external heating accompanies long-lived twists of the 
magnetosphere, $\nabla\times\bB\neq 0$, which imply long-lived electric currents $\bj$. 
The currents are sustained (and gradually dissipated) through continual electric 
discharge that fills the active j-bundle with relativistic $e^\pm$ pairs, and some of 
these particles bombard the footprint of the j-bundle. 
Figure~\ref{fig:spots} shows observational evidence for this mechanism in 
transient magnetars --- the shrinking hots spots predicted by electrodynamics
of  resistive ``untwisting'' \citep{2009ApJ...703.1044B}.
Similar localized heating is expected to operate in persistent magnetars, however, 
it appears insufficient to explain emission with large surface area $A>10^{12}$~cm$^2$.

A related puzzle of persistent magnetars is 
that their magnetospheres stay twisted much longer than in transient magnetars.
In particular, AXP~1E~1841$-$045 has been producing approximately steady 
nonthermal emission for at least one decade. Its phase-resolved hard X-ray spectrum 
is well reproduced by the model of $e^\pm$ flow in the j-bundle, and 
the soft X-ray component may be described as two blackbodies --- the warm stellar 
surface + the hot j-bundle footprint \citep{2015ApJ...807...93A}. 
At the same time, the nonthermal luminosity implies a short timescale for 
ohmic dissipation of the magnetospheric twist $t_{\rm diss} \approx 0.1\, \psi^2$~yr, 
which can hardly exceed $\sim 1$~yr (here $\psi\simlt \pi$~radian is the twist amplitude).
To survive a decade, this configuration requires energy supply from the star, 
and it is unclear how the system finds a steady state.

If the magnetar surface is indeed heated by the magnetospheric activity (through 
damping of Alfv\'en waves or particle bombardment) this still relies on 
a primary driver inside the star, regardless of how dissipative or quiet 
it may be. In particular, sustaining the magnetospheric twists 
against ohmic decay requires continual (or frequent) shear motions of the crust, 
which must be driven by the internal fields. The ultimate energy source  for both  
magnetospheric emission and surface glow must be the 
magnetic energy stored inside the star. 

Two processes can build up internal stresses that drive crustal motions: Hall drift in 
the crust and ambipolar diffusion in the core. Both, however, have their limitations. 
The Hall driver obeys a strong  upper limit given by \Eq~(\ref{eq:sH}). Hall drift can 
generate significant transient shear \citep{2016arXiv160604895L} 
but not the persistent activity with luminosity exceeding $10^{35}$~erg~s$^{-1}$.
Ambipolar diffusion naturally creates stresses at the bottom of the crust and can force it 
to flow, allowing the helical field in the core to unwind \citep{2002ApJ...574..332T}. 
The limitation here is the net flow/unwinding angle 
$\Delta s\sim (B_\phi/B_r)_{\rm core}<10^2$. 
The unwinding motion with $\Delta s\sim 10^2$ could
sustain the magnetospheric activity for only $\sim 10^2$~yr,
if it occurs with the optimal rate $\dot{s}\sim 1$~rad~yr$^{-1}$ that is just sufficient to 
offset ohmic decay of the magnetospheric twist. 
The external activity would last longer if the internal field has many twisted
domains that unwind at different times, creating a kind of a firework with the overall duration
longer than the output of each individual domain. This could perhaps bring the time-span 
of activity to the observed magnetar ages of $\sim 10$~kyr.

There is some observational support for the intermittency of the magnetic flux emerging 
from magnetars, consistent with the picture of patches of concentrated flux (flux tubes).
Evidence for an active flux tube with a magnetic field much stronger than the 
average (dipole) field was found in SGR~0418+5729 \citep{2013Natur.500..312T}.

\acknowledgements
We thank Ashley Bransgrove and the referee for useful comments on the manuscript.
This work was supported by NASA grant NNX13AI34G
and a grant from the Simons Foundation (\#446228, Andrei Beloborodov).


\begin{appendix}

\section{Approximate model for ambipolar diffusion} \label{app}

\begin{figure}[]
\begin{tabular}{c}
\includegraphics[width=0.47\textwidth]{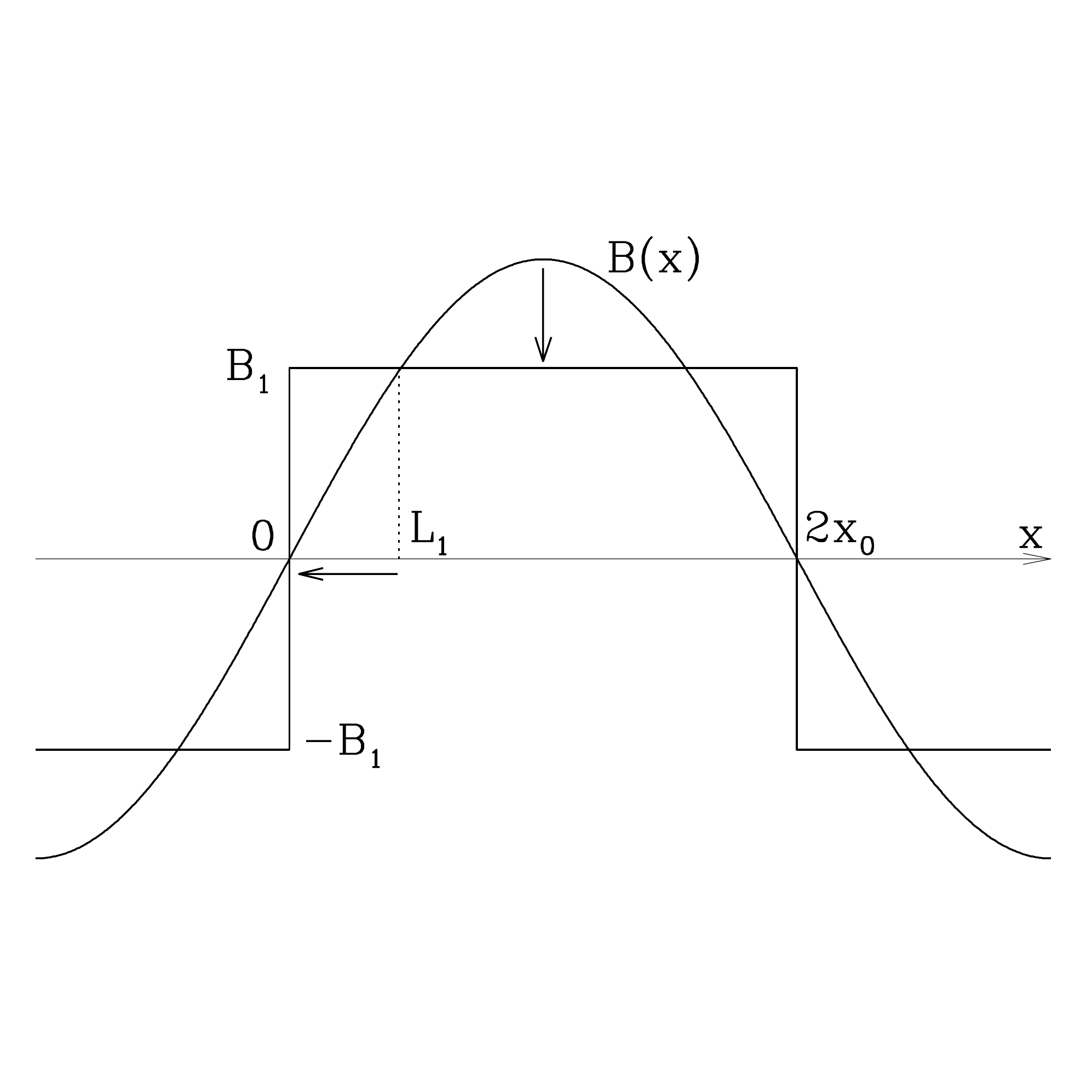}
\includegraphics[width=0.47\textwidth]{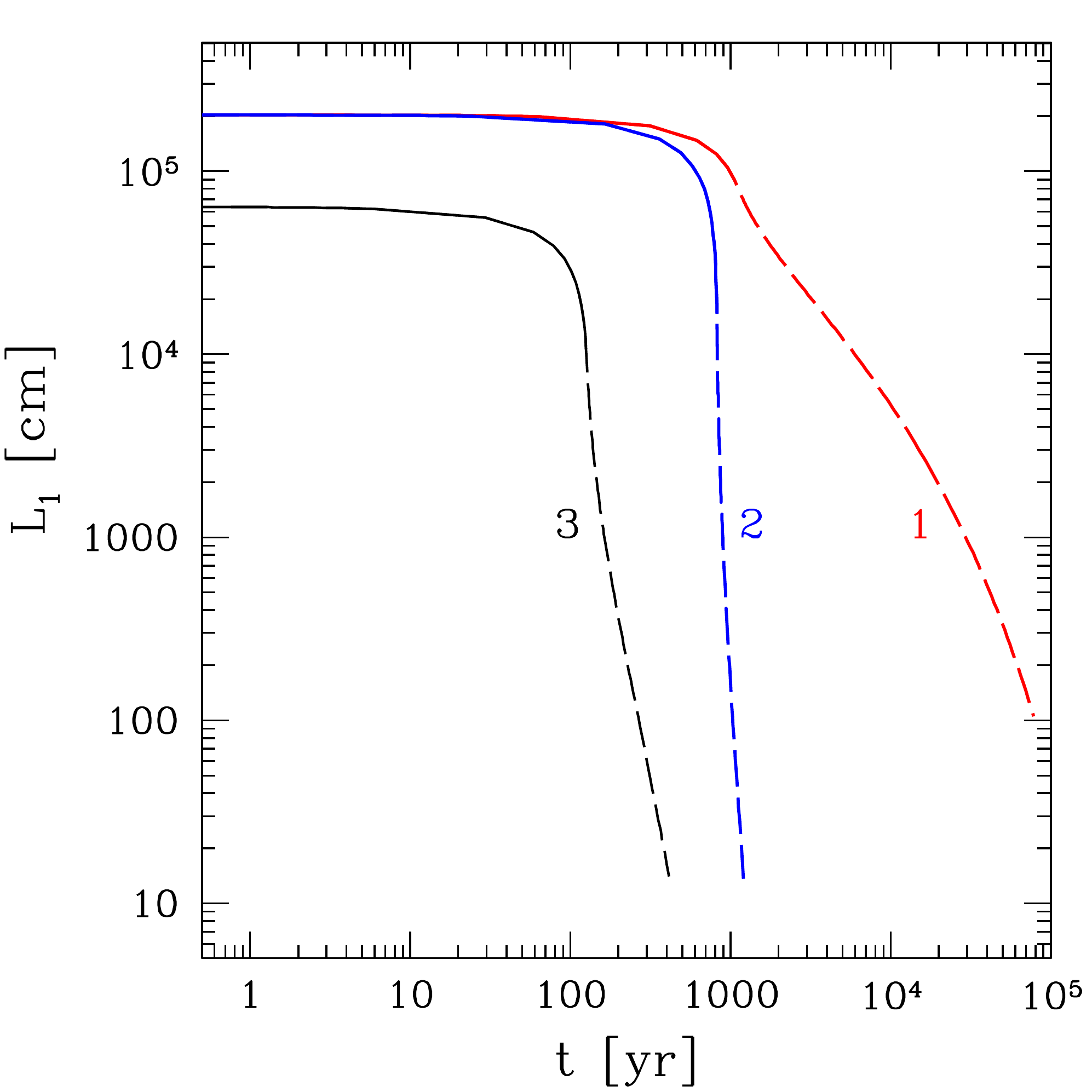}
\end{tabular}
\caption{Left: the profile of the magnetic field $B(x)$, in its initial and 
final states. The evolution is indicated by the two arrows: $L_1$ shrinks, making 
the profile steeper near the null points, while the maximum $B$ decreases, 
making the profile flatter between the null points. 
The resulting final state is close to a step function, with a steep jump of $B$ 
near the null point, which is supported by a thin current sheet.
Right: Evolution of the 
current sheet half-thickness $L_1$ in the three models shown in Figure~\ref{fig:Tc}.
Solid part of the curve shows the friction-dominated stage and dashed part 
shows the pillow stage. The moment of the hydrostatic pillow formation near the 
null point corresponds to the peak in temperature seen in Figure~\ref{fig:Tc};
the shrinking of $L_1$ is fastest at this moment.
The curves end when $L_1$ reaches $L_{\min}$ estimated in \Eq~(\ref{eq:Lmin}).
 }
 \label{fig:L1}
 \end{figure}

The one-dimensional model with the initial magnetic field $B(x)=B_0\sin(\kk x)$ 
is illustrated in the left panel of Figure~\ref{fig:L1}.
The region $0<x<\Leff$ is shrinking with rate $\dot{L}_1$ that is twice the 
local plasma speed $v_1=v(L_1)$. Note that the magnetic flux in this region
$\Psi_1=\int B\,dx=B_1L_1/2$ is decreasing, which is only possible if the boundary 
$L_1$ moves faster than the plasma. The flux transport across the boundary $L_1$
is described by $\dot{\Psi}_1=(-v_1+\dot{L}_1) B_1$, which gives 
\beq 
   \dot{L}_1=2v_1.
\eeq

As long as the plasma speed $v$ is regulated by the p-n friction (as in 
\Eq~(\ref{eq:vfric})), one finds 
\beq
  v_1=-\frac{\tpn B_1^2}{4\pi\rho_p\Leff},  \qquad 
  \frac{d\Leff}{dt}=-\frac{\tpn B_1^2}{2\pi\rho_p\Leff}, \qquad {\rm (friction~dominated)}
\eeq
which would lead to the singularity $L_1\rightarrow 0$ in a finite time.
This model is, however, incomplete, because it neglects the build up of 
pressure near the null point, which can slow down the compression.
The pressure gradient remains negligible as long as Murca reactions sufficiently 
quickly convert electrons and protons to 
neutrons (which can flow out of the compressed region across the magnetic field).
Eventually this approximation breaks and the finite rate of Murca reactions 
becomes an important limitation near the null point. This occurs when $\Leff$ 
becomes smaller than the scale $a$ given in \Eq~(\ref{eq:a}). Then a hydrostatic 
pressure ``pillow'' is formed at $x=0$ which nearly offsets the surrounding 
magnetic pressure $B_1^2/8\pi$,
\beq
   n_e\dmu\sim \frac{B_1^2}{8\pi}, \qquad L_1\ll a.
\eeq
\Eq~(\ref{eq:compr}) now yields the following compression rate near the null point,
\beq
   \frac{\partial}{\partial x} (n_ev)\approx -\frac{\lambda B_1^2}{8\pi n_e},
    \qquad \Leff\ll a.
\eeq
This gives $v(x)=v_1\,x/L_1$ with $v_1=-\lambda B_1^2 L_1/8\pi n_e^2$.
In summary,  the compression rate of the current sheet $\dot{L}_1=2v_1$ is 
controlled by p-n friction as long as $L_1\gg a$ and by Murca reactions in the 
pillow when $L_1\ll a$. \Eq~(\ref{eq:Ldot}) summarizes the two 
regimes; the transition between them, $L_1=L_\star$, 
is defined by matching the two formulas for $v_1$.

Both p-n friction and the Murca rate depend on temperature,
whose evolution is controlled by heating due to magnetic energy dissipation.
An approximate equation for magnetic dissipation may be derived as follows.
Consider the domain $0<x<x_0=\pi/2\kk$ with the (conserved) total magnetic flux, 
\beq
   \Psi_0=\int_0^{x_0} B\,dx=\frac{B_0}{\kk}=B_1 x_0=const,
\eeq
and the (decreasing) magnetic energy
\beq
   E(t)=\int_0^{x_0}\frac{B^2}{8\pi}\,dx.
\eeq
We divide the domain into two parts:
\\
(1) In the current sheet $0<x<L_1$, we use the approximation $B(x)=B_1 x/L_1$.
The magnetic flux and energy of this region are given by
\beq
 \Psi_1\approx \frac{B_1L_1}{2}, \qquad  E_1\approx \frac{B_1^2L_1}{24\pi}. 
\eeq
(2) In the region $L_1<x<x_0$, the magnetic flux $\Psi_2$ is 
\beq
  \Psi_2=\Psi_0-\Psi_1=B_1x_0-\frac{B_1L_1}{2}.
\eeq 
A simple expression for the magnetic energy of this region is found
in the linear order of $B-B_1\ll B_1$, neglecting $(B-B_1)^2$,
\beq
   E_2=\int_{L_1}^{x_0} \frac{B^2}{8\pi}\,dx
   \approx \frac{B_1\Psi_2}{4\pi}-(x_0-L_1)\frac{B_1^2}{8\pi}=\frac{x_0 B_1^2}{8\pi}.
\eeq
It remains constant and equals the final energy of the entire domain $E_{\rm fin}$.

The total magnetic energy is then given by
\beq
\label{eq:energy}
   E=E_1+E_2\approx \frac{B_1^2 L_1}{24\pi}+E_{\rm fin}.
\eeq
This equation should provide a good approximation to the magnetic energy
when $\kk L_1\ll 1$. The initial state $B(x)=B_0\sin\kk x$ has a large $L_1=2/\pi\kk$;
in this case, our approximation underestimates the energy available for dissipation, 
$E-E_{\rm fin}$, by a factor of 2.
Using the approximate  relation between $L_1$ and magnetic energy provided by 
\Eq~(\ref{eq:energy}) one finds the volume-averaged heating rate $\dqh=-\dot{E}/x_0$,
which we use in \Eq~(\ref{eq:Tdot}).

Inside the pillow ($x=0$), a significant $\dmu$ is built up, 
\beq
\label{eq:xi}
   \xi=\frac{\dmu}{kT}\sim \frac{B_1^2}{8\pi n_e kT}
   \approx 2.9\, B_{1,16}^2\, T_9^{-1} \left(\frac{\rho}{\rhonuc}\right)^{-2}.
\eeq 
Therefore, $\lambda$ in \Eq~(\ref{eq:Ldot}) must be evaluated using 
the correction factor $H(\xi)$ (see \Eq~(\ref{eq:lambda}) and \citet{1995ApJ...442..749R}),
\beq
\label{eq:H}
   H(\xi)=1+\frac{189\,\xi^2}{367\pi^2}+\frac{21\,\xi^4}{367\pi^4}+\frac{3\,\xi^6}{1835\,\pi^6}.
\eeq
Note also that $\lambda$ and $\dqnu$ are related, since both depend on the rate 
of Murca reactions. This relation is given by \citep{2001PhR...354....1Y},
\beq
\label{eq:dqnu_lam}
   \lambda_0=\frac{\lambda}{H(\xi)}=\frac{14680}{11513}\,\frac{\dqnu^0}{(\pi kT)^2},
\eeq
where $\dqnu^0$ is the Murca cooling rate at $\dmu\ll kT$, and 
$\lambda_0$ describes the rate of $\dmu$ relaxation for $\dmu\ll kT$. 
We use $\dqnu=\dqnu^0$, because most of neutrino losses occur in the region 
$x>L_1$ where $\dmu$ remains small. 

During the main heating stage there is an approximate balance between heating 
and cooling $\dqh\approx\dqnu$, which gives
\begin{eqnarray}
\label{eq:TL1}
  L_1\approx \left\{\begin{array}{ll}
\vspace*{0.2cm}
      \displaystyle{ \frac{\tpn B_1^4\,\kk}{24\pi^3\rho_p\, \dqnu}} & \quad L_1>\Ls, \\
      \displaystyle{\frac{12}{\xi^2H(\xi)\,\kk}} & \quad L_1<\Ls. 
                           \end{array}
                   \right.
\end{eqnarray}
This provides a relation between $T$ and $L_1$, and then it is sufficient to solve one 
differential equation, e.g. \Eq~(\ref{eq:Tdot}) for $T(t)$. In particular, the 
transition $L_1=\Ls$ typically occurs in the regime $\dqh\approx\dqnu$. 
One can solve for $\xi$ and $T$ at the transition by matching the two expressions 
in \Eq~(\ref{eq:TL1}) and using \Eq~(\ref{eq:xi}),
\beq
\label{eq:xis}
   \xi_\star\approx 4\;\kk_{-5}^{-1/6}\,B_{1,16}^{4/3}\,\left(\frac{\rho}{\rhonuc}\right)^{-3/2},
   \qquad  T_\star\approx  7.2\times 10^8\,\kk_{-5}^{1/6}\, B_{1,16}^{2/3} 
    \left(\frac{\rho}{\rhonuc}\right)^{-1/2} {\rm K},
\eeq
where we have used the approximation $\xi H^{1/12}\approx \xi$.
A significant deviation from the balance 
$\dqh\approx\dqnu$
develops at later stages; then \Eq~(\ref{eq:TL1}) becomes invalid and the evolution is 
found from the coupled differential equations for $T(t)$  and $L_1(t)$.

Figure~\ref{fig:L1} shows the evolution of $L_1(t)$ for the sample models presented
in Figure~\ref{fig:Tc}.
The initial evolution on the friction timescale takes less than 1~kyr,  then the pillow 
forms, however it does not stop the fast collapse of the current sheet.
The compression timescale $L_1/|\dot{L}_1|$ is then controlled by Murca reactions,
\beq
\label{eq:tl}
  t_\lambda=\frac{4\pi n_e^2}{\lambda B_1^2}
  \approx \frac{80 {\rm~yr}}{(B_{1,16})^2\, T_9^6\, H(\xi)}\,\left(\frac{\rho}{\rhonuc}\right)^{10/3}.
\eeq
An upper limit to this timescale is obtained if $\xi\gg 10$; then 
$H(\xi)\approx (0.11\xi)^6$ and 
$\tlam\approx 80\,(B_{1,16})^{-14} (\rho/\rhonuc)^{46/3}$kyr.
However, before the regime $\xi\gg 10$ is approached, the effects of 
a finite electric conductivity become important and stop the shrinking of $L_1$.
The effective conductivity (associated with ohmic dissipation) across
the magnetic field $\bB$ is approximately equal to the conductivity along $\bB$
(see \Sect~\ref{ohm}), which in the core is given by 
$\cond\approx 4.2\times 10^{26}\, T_9^{-2}(\rho/\rhonuc)^3$~s$^{-1}$ \citep{1990A&A...229..133H}.
Magnetic diffusivity $\eta=c^2/4\pi\cond$ stops the compression of $L_1$ when 
$L_1v_1\sim c^2/4\pi\cond$. This gives the minimum thickness of the current sheet,
\beq
\label{eq:Lmin}
   L_{\min}\approx 1\,T_9  \left(\frac{\rho}{\rhonuc}\right)^{-3/2}
        \left(\frac{t_\lambda}{1{\rm ~kyr}}\right)^{1/2}  {\rm m}.
\eeq

\end{appendix}

\bibliography{ms}

\end{document}